\documentclass[aps,pre,twocolumn,groupedaddress,showpacs]{revtex4-2}
\newcommand{\bec}[1]{\mbox{\boldmath $ #1$}}
\usepackage{bm}
\usepackage{graphicx}
\usepackage{color}

\newcommand{\meanrho}{\overline{\rho}}

{}
\newcommand{\meanUU}{\overline{\bm{U}}}

{}
{}

\newcommand{\meanN}{\overline{n}}

\newcommand{\meanP}{\overline{P}}
\newcommand{\meanT}{\overline{T}}

\newcommand{\meanU}{\overline{U}}
\newcommand{\meanS}{\overline{S}}

\begin{document}
\title{Experimental study of turbulent thermal diffusion of inertial particles
in a convective turbulence forced by oscillating grids}
\author{E.~Elmakies}
\author{O.~Shildkrot}
\author{N. Kleeorin}
\author{A. Levy}
\author{I.~Rogachevskii}
\email{gary@bgu.ac.il}

\bigskip
\affiliation{
The Pearlstone Center for Aeronautical Engineering
Studies, Department of Mechanical Engineering,
Ben-Gurion University of the Negev, P.O.Box 653,
Beer-Sheva 8410530,  Israel}

\date{\today}
\begin{abstract}
We investigate the phenomenon of  turbulent
thermal diffusion of inertial solid particles in laboratory experiments with
convective turbulence forced by one or two oscillating grids in the air.
Turbulent thermal diffusion causes a non-diffusive contribution to turbulent
flux of particles described in terms of an effective drift velocity
directed opposite to the gradient of the mean fluid temperature.
For inertial particles, this effective drift velocity
depends on the Stokes and Reynolds numbers.
In the experiments, fluid velocity and spatial distribution of
inertial particles are measured using a Particle Image Velocimetry (PIV) system,
and the temperature field is measured in many locations by
a temperature probe equipped with 12 thermocouples.
Measurements of temperature and particle number density spatial distributions
have demonstrated the formation of large-scale clusters of inertial particles
in the vicinity of the mean temperature minimum due to turbulent thermal diffusion.
In the experiments, the effective drift velocity caused by turbulent thermal diffusion that results in the formation of large-scale clusters
of inertial particles (having the diameter $10 \mu m$)
is in 1.5 -- 2.5 times larger than that for noninertial particles (having the diameter $0.7 \mu m$)
depending on the level of turbulence.
This is in agreement with the theoretical predictions.
\end{abstract}

\maketitle

\section{Introduction}
\label{sect1}

Turbulent transport of particles
has been investigated more than a hundred years
due to a number of applications
in astrophysical, geophysical and industrial flows
\cite{CSA80,ZRS90, BLA97, SP06, ZA08,CST11,RI21}.
However, some key questions are still the subject
of ongoing discussions.

One of the key questions is related to  {\it mechanisms
of formation of large-scale inhomogeneous distributions of small particles},
which are called as {\it large-scale particle clusters} and are
caused by {\it small-scale turbulence}.
The spatial scales of such particle clusters are much larger than the
integral turbulence scale and their lifetimes are much longer than the turbulent time \cite{RI21}.
Examples of the key phenomena resulting in the formation of large-scale particle clusters are turbophoresis
of inertial particles in a small-scale inhomogeneous turbulence \cite{CTT75,RE83}
and turbulent thermal diffusion in a small-scale temperature-stratified turbulence
\cite{EKR96,EKR97}.

Turbophoresis is interpreted in terms of an effective drift velocity
${\bm V}_{\rm turb} = - \kappa_{\rm turb} \, {\bm \nabla} \langle{\bm u}^2\rangle$
of inertial particles,
where ${\bm u}$ are
fluid velocity fluctuations and $\kappa_{\rm turb}$ is the turbophoretic
coefficient which depends on the Stokes and Reynolds numbers
\cite{CTT75,RE83,G97,EKR98,G08,SSB12,LCB16,MHR18,KHB95,RR04}.
The mechanism of turbophoresis of inertial particles is related to an effect of particle inertia
advected by a non-stratified inhomogeneous turbulence.
The inertia results in a local drift of particles by centrifugal force
to the external regions between turbulent eddies.
Since there is a preferential direction
due to a non-zero large-scale gradient of the turbulence intensity
$\bec{\nabla}\left\langle{\bm u}^2\right\rangle$ in a small-scale inhomogeneous turbulence,
the total effect of the particle drift to the regions with low turbulence intensity
does not vanish.
Due to turbophoresis, inertial particles are accumulated in the vicinity of the minimum of the turbulent intensity.

Another phenomenon related to formation of large-scale particle clusters is
turbulent thermal diffusion that causes a nondiffusive turbulent flux of
particles in the direction of the turbulent heat flux and results in
formation of large-scale inhomogeneous distributions of particles
in the vicinity of mean temperature minimum in a small-scale
temperature-stratified turbulence.
This effect originates from
the turbulent flux of particles $\left\langle {\bm u}  \, n'  \right\rangle
= {\bm V}^{\rm eff} \, \meanN - D_{\rm T} \, {\bm \nabla} \meanN$,
where ${\bm V}^{\rm eff} =- \alpha \, D_{\rm T} \, {\bm \nabla} \ln \meanT$ is the effective drift velocity
caused by turbulent thermal diffusion \cite{EKR96,EKR97,RI21},
$\meanN$ is the mean particle number density,
$\meanT$ is the mean temperature of the fluid,
$n' $ are fluctuations of particle number density
and $D_{\rm T}$ is the turbulent diffusion coefficient.
For noninertial particles the parameter $\alpha=1$,
while for inertial particles the parameter $\alpha> 1$
and it depends on the Stokes and Reynolds numbers
\cite{RI21,EKR96,EKR97,EKR98} (see Sec. ~\ref{sect2},
where the physics related to the phenomenon
of turbulent thermal diffusion is discussed).

We stress that turbulent thermal diffusion and turbophoresis
are two principally different phenomena.
Turbulent thermal diffusion of noninertial and inertial particles
is a purely collective phenomenon occurring in
temperature-stratified turbulence and originating from the turbulent particle flux (i.e.,
the second-order correlation of velocity--number density fluctuations), and it
depends on the turbulent heat flux (i.e., the second-order correlation of velocity--temperature fluctuations).
In contrast to turbulent thermal diffusion, phenomenon of turbophoresis of inertial particles is originated from the averaging
of particle velocity field and intrinsically related to particle inertia
in inhomogeneous turbulence.

Turbulent thermal diffusion for noninertial and inertial particles
has been investigated analytically adopting various theoretical approaches
\cite{RI21,EKR96,EKR97,EKR98,EKR00,EKRS00,EKRS01,PM02,RE05,AEKR17}.
This phenomenon has been detected in direct numerical simulations (DNS) \cite{HKRB12,RKB18}
in a forced stably-stratified turbulence.
Analytical and numerical studies have shown that particles are accumulated
inside large-scale clusters in the vicinity of the mean temperature minimum
due to phenomenon of turbulent thermal diffusion.

Turbulent thermal diffusion has been applied in geophysical \cite{SSEKR09}, planetary \cite{EKPR97}
and astrophysical  \cite{H16,KR25} turbulence.
It was found \cite{SSEKR09} that turbulent thermal diffusion results in formation of long-living large-scale aerosol
layers in the vicinity of temperature inversions in the tropopause, and
the predictions based on the theory of turbulent thermal diffusion
are in a good agreement with the observed profiles of large-scale
aerosol concentration and mean temperature in the vicinity of the tropopause of the Earth atmosphere.
Turbulent thermal diffusion  explains the formation of planetesimals (progenitors of planets) in protoplanetary
disks \cite{H16,KR25} and the formation of aerosol concentrations in the upper atmosphere of Titan \cite{EKPR97}.
This phenomenon is also important in environmental and engineering systems \cite{RI21}.

Turbulent thermal diffusion has been
studied in various  {\it laboratory experiments only for noninertial particles}.
In particular, this effect has been detected for noninertial particles
in laboratory experiments  in the airflow
in temperature stratified turbulence
produced by one or two oscillating grids
\cite{BEE04,EEKR04,EEKR06a,AEKR17,EKRL22,EKRL23},
by multi-fan turbulence generator \cite{EEKR06b}
and in strongly inhomogeneous and anisotropic convection
forced by two similar turbulence generators with oscillating membrane
and a steady grid \cite{ZEKRL25}.
This phenomenon has been even found for nanoparticles \cite{SKRL22}
in turbulent convection.

However, turbulent thermal diffusion has not yet been
investigated for {\it inertial particles in laboratory experiments}.
Since most systems contain inertial particles, this study has a crucial importance.
In the present paper, we investigate phenomenon of  turbulent
thermal diffusion of {\it inertial solid particles} in laboratory experiments with
a convective turbulence forced by one or two oscillating grids in the airflow.
We present a systematic comparison between the behaviors of inertial and noninertial particles
in a small-scale temperature-stratified turbulence produced in the laboratory experiments.

This paper is organized as follows.
In Sec. ~\ref{sect2} we discuss the theoretical background related to turbulent thermal diffusion.
In Sec. ~\ref{sect3} we describe experimental setup
and measurement techniques.
In Sec. ~\ref{sect4} we analyse the obtained experimental results, and
in Sec.~\ref{sect5} we outline conclusions.

\section{Turbulent thermal diffusion}
\label{sect2}

In this section we discuss the theoretical background related to turbulent thermal diffusion.
We first consider small noninertial particles in a turbulent fluid which move with the fluid velocity.
Dynamics of the particle number density $n(t,{\bm x})$
in a low-Mach-number fluid velocity field ${\bm U}(t,{\bm x})$ is governed by the
convective-diffusion equation
\begin{eqnarray}
{\partial n \over \partial t} + {\bm \nabla} {\bf \cdot}({\bm U}\, n) = D \, \Delta n ,
\label{MD1}
\end{eqnarray}
where
$D= k_B \,T/(3\pi \rho \, \nu \, d_{\rm p})$ is the coefficient
of the molecular (Brownian) diffusion of particles,   $d_{\rm p}$ is the particle diameter,
$\nu$  is the kinematic viscosity of the fluid, $T$ and $\rho$  are the fluid temperature and
density,  $k_B$ is the Boltzmann constant, Ma $=|{\bm U}|/c_{\rm s} \ll 1$ is the Mach-number
and $c_{\rm s}$ is the sound speed.
For a low-Mach-number, the continuity equation for the fluid density
is applied in an anelastic approximation $\bec{\nabla} {\bf \cdot}(\rho \, {\bm U}) = 0 $.

We investigate dynamics of  particles in spatial scales which are much larger than the integral scale of turbulence $\ell_0$, and in the time scales which are much longer than the turbulent time scales $\tau_0$.
We apply a mean-field approach, whereby all quantities are decomposed into the
mean and fluctuating fields, and use the Reynolds averaging.
In particular, the particle number density $n= \overline{n} + n'$, where $\overline{n}=\langle n \rangle$ is the mean particle number density, $n'$ are particle number density fluctuations and $\langle n' \rangle=0$. The angular brackets denote ensemble averaging. Averaging Eq.~(\ref{MD1}) over an ensemble of turbulent velocity field, we obtain the mean-field equation for the particle number density:
\begin{eqnarray}
{\partial \overline{n} \over \partial t} + {\bm \nabla} {\bf \cdot} \left(\overline{\bm U} \, \overline{n} + \langle {\bm u} \, n'  \rangle \right)= D \, \Delta \overline{n} ,
\label{MD2}
\end{eqnarray}
where $\langle {\bm u} \, n' \rangle$ is the turbulent flux of particles
and ${\bm u}$ are fluid velocity fluctuations.

To determine the particle turbulent flux $\langle {\bm u} n' \rangle$, we use an equation for particle number density fluctuations $n'$ obtained by subtracting Eq.~(\ref{MD2}) from Eq.~(\ref{MD1}):
\begin{eqnarray}
{\partial n' \over \partial t} + {\bm \nabla} {\bf \cdot} \left({\bm u} \, n'  - \langle {\bm u} \, n'  \rangle \right) - D  \Delta n' = -({\bm u} {\bf \cdot} {\bm \nabla}) \overline{n} -\overline{n} ({\bm \nabla} {\bf \cdot} {\bm u}) ,
\nonumber\\
\label{MD3}
\end{eqnarray}
where ${\bm \nabla}{\bf \cdot} \left({\bm u} \, n'  - \langle {\bm u} \, n' \rangle \right)$ are the nonlinear terms, and
$-({\bm u} {\bf \cdot} {\bm \nabla}) \overline{n} -\overline{n} ({\bm \nabla} {\bf \cdot} {\bm u})$ are the source terms for particle number density fluctuations.
In the frame of the mean-field approach, the fluid density $\rho= \overline{\rho}  + \rho'$ is decomposed into the mean fluid density $\overline{\rho}$ and fluctuations $\rho'$, and for low-Mach-number flows $|\rho'| \ll \overline{\rho}$.
The anelastic approximation implies that
${\bm \nabla} {\bf \cdot} {\bm u} = {\bm u} \cdot {\bm \lambda}$, where
${\bm \lambda} = - {\bm \nabla} \overline{\rho} / \overline{\rho}$.
The ratio of the nonlinear term to the diffusion term in Eq.~(\ref{MD3}) is the P\'{e}clet number for particles,
which is estimated as ${\rm Pe} = u_0 \, \ell_0 / D$, where $u_0$ is the characteristic turbulent velocity
in the integral scale $\ell_0$ of turbulence.
When the spatial density of particles $n \, m_p$ is much smaller than the fluid density $\rho$,
there is a one-way coupling which implies that we
take into account the effect of the turbulent velocity on the particle number density, but
neglect the feedback effect of the particle number density on the turbulent fluid flow. Here $m_p$ is the particle mass.

For simplicity, we apply the dimensional analysis to solve Eq.~(\ref{MD3}). The dimension of
the left-hand side of Eq.~(\ref{MD3}) is the rate of change of particle number density fluctuations $n' / \tau_{n'}$, where $\tau_{n'}$ is the characteristic time of particle number density fluctuations.
For large Reynolds and P\'{e}clet numbers, the characteristic time of particle number density fluctuations $\tau_{n'}$
can be identified with the correlation time $\tau_0$ of the turbulent velocity field.
In the framework of the dimensional analysis, we replace the left-hand side of Eq.~(\ref{MD3}) by $n' / \tau_0$, and obtain
the solution  of Eq.~(\ref{MD3}) as
$n' = - \tau_0 \, \left[({\bm u} {\bf \cdot} {\bm \nabla}) \overline{n} + \overline{n} \, ({\bm \nabla} {\bf \cdot} {\bm u})\right] $.
Multiply this equation by $u_i$ and average it
over an ensemble of turbulent velocity field, we obtain the turbulent flux of particles as
\begin{eqnarray}
\left\langle u_i  \, n' \right\rangle &=& - \tau_0 \,\left\langle u_i u_j  \right\rangle \, \nabla_j \overline{n} - \tau_0 \, \overline{n} \,\left\langle u_i (\bec{\nabla} {\bf \cdot} {\bm u}) \right\rangle
\nonumber\\
&\equiv& V_i^{\rm eff} \, \overline{n} - D_{ij}^{(n)} \, \nabla_j \overline{n} ,
\label{MD5}
\end{eqnarray}
where $D_{ij}^{(n)} = \tau_0 \,\left\langle u_i u_j  \right\rangle$
is the turbulent diffusion tensor.
For simplicity, we consider an isotropic turbulence, so that
$\langle u_i u_j  \rangle = \delta_{ij} \, \langle {\bm u}^2  \rangle/3$
and the turbulent diffusion tensor for large P\'{e}clet numbers is given by
$D_{ij}^{(n)} =D_{\rm T} \delta_{ij}$, where $D_{\rm T}= \tau_0 \,\left\langle {\bm u}^2  \right\rangle / 3$
is the turbulent diffusion coefficient.

The term ${\bm V}^{\rm eff} \, \overline{n}$ in
Eq.~(\ref{MD5}) describes the contribution to the turbulent flux of particles due to
the effective drift velocity:
${\bm V}^{\rm eff} = - \tau_0 \,\left\langle {\bm u} (\bec{\nabla} {\bf \cdot} {\bm u}) \right\rangle$, which
in the anelastic approximation is given by
$V_i^{\rm eff} = - \tau_0 \, \left\langle u_i u_j \right \rangle \, \lambda_j$.
Therefore, for an isotropic turbulence, the effective drift velocity
for noninertial particles is given by
${\bm V}^{\rm eff}  = D_{\rm T} \, {\bm \nabla} \ln \overline{\rho}$,
and the particle turbulent flux $\langle  {\bm u} \, n'  \rangle$ is
\begin{eqnarray}
\left\langle {\bm u} \, n'  \right\rangle = {\bm V}^{\rm eff} \, \overline{n} - D_{\rm T} \, {\bm \nabla} \overline{n} .
\label{MD8}
\end{eqnarray}
Now we use the equation of state for a perfect gas,
$P=(k_B/ m_\mu) \, \rho \, T$, that can be also rewritten for the mean fields as
$\overline{P}=(k_B/ m_\mu) \, \overline{\rho} \, \overline{T}$,
where $\overline{P}$ and $\meanT$ are the mean pressure
and mean temperature, respectively.
Here we assume that $\overline{\rho} \, \meanT \gg \langle \rho' \, \theta \rangle$.
Applying the equation of state, we obtain
${\bm \nabla} \, \ln\overline{\rho} = {\bm \nabla} \ln \overline{P} - {\bm \nabla} \ln \meanT$.
For small mean pressure gradient, ${\bm \nabla} \, \ln\overline{\rho} \approx  - {\bm \nabla} \meanT$, so that
the effective drift velocity for noninertial particles is given by \cite{EKR96,EKR97,RI21}
\begin{eqnarray}
{\bm V}^{\rm eff} = - D_{\rm T} \, {{\bm \nabla} \meanT \over \meanT} .
\label{MD9}
\end{eqnarray}
Various analytical methods (e.g., the quasi-linear approach,
the $\tau$ approaches applied in physical and Fourier spaces,
the path integral approach and the direct interaction approximation)
yield result similar to Eq.~(\ref{MD9}), see, e.g.,
Refs.~\cite{RI21,EKR96,EKR97,EKR98,EKR00,EKRS00,EKRS01,PM02,RE05,AEKR17}.

For inertial particles, the effective drift velocity is given by
\begin{eqnarray}
{\bm V}^{\rm eff} = - \alpha \, D_{\rm T} \, {{\bm \nabla} \meanT \over \meanT} ,
\label{MD10}
\end{eqnarray}
where the parameter $\alpha$ is \cite{RI21,EKR96,EKR97,EKR98}
\begin{eqnarray}
\alpha = 1 + 2\, {V_{\rm g} \, L_P \, \ln{\rm Re} \over u_0 \, \ell_0} .
\label{A10}
\end{eqnarray}
Here ${\bm V}_{\rm g} = \tau_{\rm p} \, {\bm g}$ is the terminal fall velocity of inertial particles,
$\tau_{\rm p}= m_{\rm p} / (3 \pi \rho \, \nu d_{\rm p})$ is the Stokes time,
${\bm g}$ is the acceleration due to the gravity, $\rho$
is the fluid density, $\nu$ is the kinematic viscosity, $m_{\rm p}$ and $d_{\rm p}$ are the particle mass and diameter, respectively,
$L_P = |\nabla_z \meanP / \meanP |^{\, -1}$ is the pressure height scale, $\meanP$ is the fluid mean pressure and
${\rm Re} = u_0 \, \ell_0/ \nu$ is the Reynolds number.

\begin{figure*}[t!]
\centering
\includegraphics[width=8.5cm]{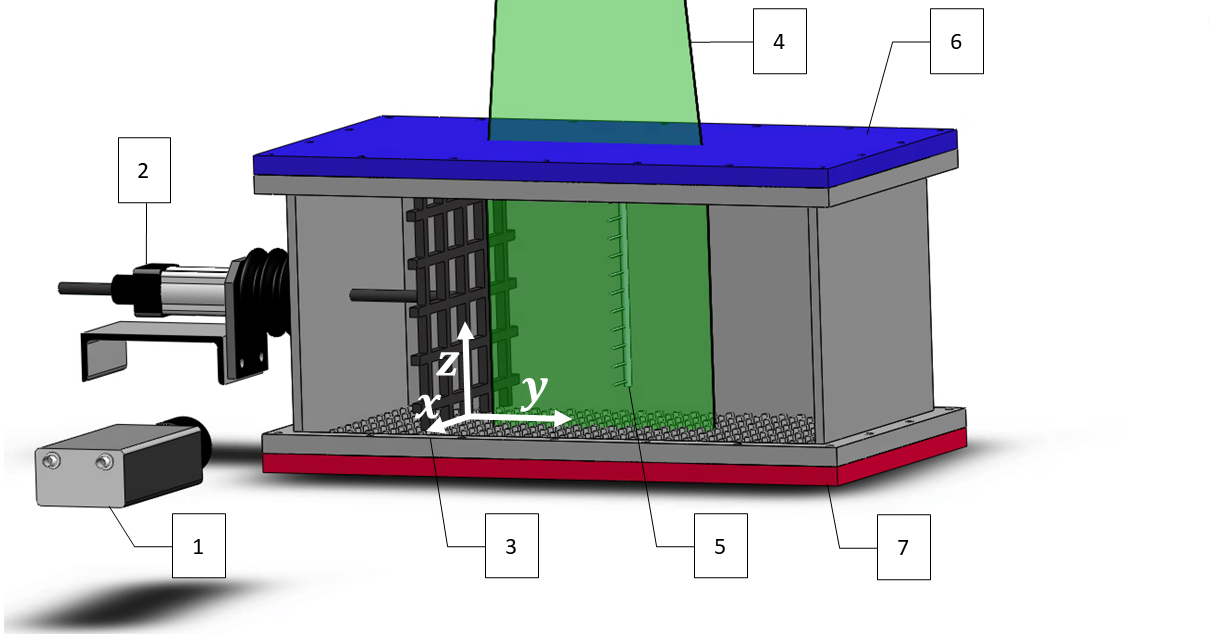}
\includegraphics[width=8.5cm]{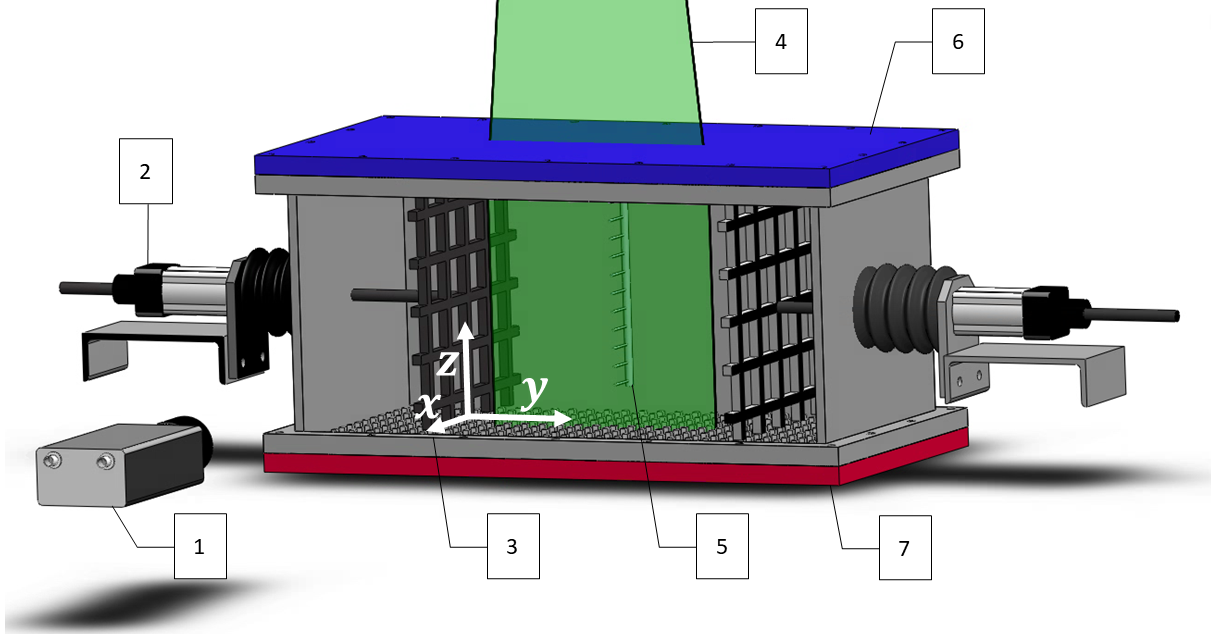}
\caption{\label{Fig1}
Experimental setup  with the convective turbulence
forced by one oscillating grid  (left panel) and by two oscillating grids (right panel):
(1) digital CCD camera; (2) rod driven by the
speed-controlled motor; (3) oscillating grid;  (4) laser light sheet;
(5) temperature probe equipped with 12 E - thermocouples;
(6) heat exchanger at the top  cooled wall of the chamber;
(7) heat exchanger at the bottom heated wall of the chamber.
}
\end{figure*}

The mean particle number density of inertial particles is determined by the following equation:
\begin{eqnarray}
{\partial \meanN \over \partial t} + {\bm \nabla} {\bf \cdot} \biggl[\biggl(\meanUU + {\bm V}_{\rm g} - \alpha \, D_{\rm T} \, {{\bm \nabla} \meanT \over \meanT}\biggr) \meanN - D_{\rm T} {\bm \nabla} \meanN\biggr]  = 0 ,
\nonumber\\
\label{A1}
\end{eqnarray}
where we neglect small molecular (Brownian) diffusion coefficient $D$ for large P\'{e}clet numbers
and take into account the terminal fall velocity ${\bm V}_{\rm g}$ of inertial particles.

For simplicity, we assume that the gradients along the vertical axis $Z$ of the mean temperature, mean velocity,
and mean number density are much larger than those in the horizontal directions.
This condition corresponds to that observed in the experiments discussed in this paper.
The steady-state solution for Eq.~(\ref{A1}) for a zero total particle flux at the boundaries
reads
\begin{eqnarray}
{\meanN(Z) \over \meanN_{\rm b}} =  \biggl({\meanT(Z) \over \meanT_{\rm b}}  \biggr)^{-\alpha} \, \exp \biggl(\int_0^Z
{\meanU_z - V_{\rm g} \over D^{^{\rm T}}_z} \,{\rm d} Z' \biggr) ,
\label{A2}
\end{eqnarray}
where $D^{^{\rm T}}_z=\ell_z \, \left[\langle u_z^2\rangle\right]^{1/2}$ is the vertical turbulent diffusion coefficient, and the boundary conditions are $\meanN_{\rm b} = \meanN(Z=0)$ and $\meanT_{\rm b}=\meanT(Z=0)$.
Equation~(\ref{A2}) implies that maximum particle number density is attained in the region
of minimum mean fluid temperature.
Therefore, turbulent thermal diffusion causes formation of large-scale particle clusters
in the vicinity of the mean temperature minimum.

The mechanism for turbulent thermal diffusion for inertial particles
is as follows.
The inertia causes particles inside the turbulent eddies to drift
out to the boundary regions between eddies due to the centrifugal inertial force.
These regions have low vorticity, maximum fluid pressure fluctuations and high
strain rate.
In particular, in regions with maximum fluid pressure fluctuations $p$ (where
$\nabla^2 p < 0)$, there is an accumulation of
inertial particles, i.e.,  ${\rm d} n' / {\rm d}t \propto - \overline{n} \, {\bm \nabla} \cdot \, {\bm u}^{\rm(p)}
= - \overline{n} \, (\tau_{\rm p} /\overline{\rho}) \,\bec\nabla^2 p
> 0$.
Here we take into account that velocity fluctuations ${\bm u}^{\rm(p)}$ of inertial particles
are not divergence free even for incompressible fluid flow, i.e.,
${\bm \nabla} \cdot \, {\bm u}^{\rm(p)} = (\tau_{\rm p} /\overline{\rho})  \,\bec\nabla^2 p + {\rm O}(\tau_{\rm p}^2/\tau_0^2)$
\cite{M87,EKR96b,RI21}, where the Stokes time $\tau_{\rm p}$ is much smaller than turbulent correlation time
$\tau_0$.
Similarly, there is an outflow of inertial
particles from regions with minimum fluid
pressure.

In homogeneous and isotropic turbulence with a zero gradient
of the mean temperature, there is no preferential direction,
so that there is no large-scale effect of particle accumulation,
and the pressure (temperature) of
the surrounding fluid is not correlated with the turbulent velocity field. The only
non-zero correlation is $\langle({\bm u} \cdot {\bm \nabla})p\rangle$,
which contributes to the flux of the turbulent
kinetic energy density.

In a temperature-stratified turbulence, fluctuations of fluid temperature $\theta$
and fluid velocity ${\bm u}$ are correlated due to a
non-zero turbulent heat flux, $\langle
\theta \, {\bm u} \rangle\not=\bm{0}$. Fluctuations of
temperature cause pressure fluctuations, which
result in fluctuations in the number density of
particles.
Increase of the fluid pressure fluctuations
is accompanied by an accumulation of particles,
and the direction of the turbulent flux of particles
coincides with that of the turbulent heat flux.
The turbulent flux of particles is directed toward the
minimum of the mean temperature, and the
particles tend to be accumulated in this
region.

To demonstrate that the directions of the mean
flux of particles and the turbulent heat flux
coincide, we assume that the mean temperature
$\overline{T}_2$ at point $2$ is larger than the mean
temperature $\overline{T}_1$ at point $1$.
We consider two small control volumes {``a''} and
{``b''} located between these two points.
Let the direction of the local turbulent velocity in
volume {``a''} at some instant be
the same as the direction of the turbulent heat
flux $\langle \theta \, {\bm u} \rangle$ (i.e., along the
$x$--axis toward point $1$) and let the local
turbulent velocity in volume
{``b''}, at the same instant, be directed
opposite to the turbulent heat flux (i.e., toward
point $2$).

In a temperature-stratified turbulence
with a non-zero turbulent heat flux $\langle \theta \, {\bm u} \rangle$,
fluctuations of the fluid pressure $p$ and velocity ${\bm u}$
are correlated, and regions with a higher level
of pressure fluctuations have higher temperature
and velocity fluctuations.
Fluctuations of temperature $\theta$ and pressure $p$ in
volume {``a''} are positive because $\theta \, u_{x} > 0$, and negative in volume
{``b''}.
Fluctuations of particle number density $n'$ are
positive in volume {``a''} (because inertial
particles are locally accumulated in the vicinity
of the maximum of pressure fluctuations, and they are negative in
volume {``b''} (because there is an
outflow of particles from regions with low
pressure fluctuations). The flux of particles $n' \, {\rm u}^{\rm(p)}$ is positive in volume {``a''}
(i.e., it is directed toward point $1$), and it
is also positive in volume {``b''}
(because both fluctuations of velocity and number
density of particles are negative in
volume {``b''}),
where ${\rm u}^{\rm(p)}$ is the particle velocity
along the $x$-axis.
Here we take into account that for small Stokes number, deviation of the particle velocity ${\rm u}^{\rm(p)}$
from the fluid velocity ${\rm u}$ is small.
Therefore, the turbulent flux of particles
$\langle n' \, {\bm u}^{\rm(p)} \rangle \approx \langle n' \, {\bm u} \rangle$ is directed, as is the turbulent heat
flux $\langle \theta \, {\bm u}\rangle$, toward point~1.
This causes the formation of large-scale
inhomogeneous structures in the spatial
distribution of inertial particles in the
vicinity of the mean temperature minimum.
\\

\section{Experimental setup}
\label{sect3}

In this section we describe the experimental set-up
and measurement technique.
We study the phenomenon of  turbulent
thermal diffusion of inertial solid particles in experiments with
a convective turbulence forced by one or two oscillating grids in the airflow.
We conduct experiments in rectangular transparent chamber
with dimensions $L_x \times L_y \times L_z$ with
$L_x=L_z=26$ cm and $L_y=53$ cm,
where the axis $Z$ is directed along the vertical
direction and the axis $Y$ is perpendicular to the grid plane.
The oscillating grids with bars arranged in a square array are parallel to the side walls of the chamber
(in the $XZ$ plane), and are positioned at a distance of 10 cm (two grid meshes)
from the side walls of the chamber, so that the oscillations of the grids occur along the $Y$ axis (see Fig.~\ref{Fig1}).
The frequency $f$ and the amplitude $A_g$ of the grid oscillations are $f=10.5$ Hz and $A_g=6$ cm,
respectively, which yield the maximum turbulence intensity in the experimental set-up.
Based on the oscillation frequency and amplitude, we estimate the grid Reynolds number
Re$_g=f A_g^2/\nu=2520$ to characterize the forcing conditions.

To study turbulent thermal diffusion, we use forced convection, where
the convective turbulence is produced by buoyancy due to
the temperature difference $\Delta T$
between the bottom and upper walls of the chamber.
Forcing by the oscillating grids allows to destroy the large-scale circulation
and decrease mean velocity field in comparison with velocity fluctuations.
In all experiments with turbulent convection, the temperature difference $\Delta T = 50$~K
between the bottom and upper walls of the chamber is applied.

Two aluminium heat exchangers with rectangular pins $3 \times 3 \times 15$
mm are attached to the bottom (heated) and top (cooled)
walls of the chamber,
which allow one to form a large vertical mean temperature gradient
in the core of the fluid flow.
The temperature field is measured by means of a temperature probe equipped with 12
E - thermocouples.
The thermocouples with the diameter of 0.13 mm and
the sensitivity of $\approx 75 \, \mu$V/K are attached to
a vertical rod with a diameter 4 mm, and the mean distance between thermocouples
is about 21.6 mm (see for details Ref.~\cite{EKRL22}).
The data are recorded using the developed software based
on LabView 2024 Q1, and the temperature maps are obtained
using MATLAB~R2024a.

The temperature field has been measured in many locations
using an array of thermocouples distributed along a vertical probe.
The two-dimensional temperature field is reconstructed by performing measurements
at multiple horizontal positions (every 1 cm).
The resulting discrete data points are interpolated onto a regular grid
using a linear interpolation scheme in MATLAB to obtain the continuous temperature field.
By time averaging in every point, we determine a mean
temperature field.

The velocity field is measured with a
Particle Image  Velocimetry (PIV) system \cite{AD91,RWK07,W00},
consisting of a Nd-YAG laser (Continuum Surelite $2 \times
170$ mJ) and a progressive-scan 12 bit digital CCD
camera (with pixel size $6.45 \, \mu$m $\times \,
6.45 \, \mu$m and $1376 \times 1040$ pixels).
As a tracer for the PIV measurements,
we use an incense smoke with spherical solid particles
having the mean diameter of $0.7 \mu$m
and the material density $\rho_{\rm p}\approx 10^3 \meanrho$, where
$\meanrho$ is the mean fluid density.
The particles are produced by high temperature sublimation of solid
incense grains (see for details Ref.~\cite{EKRL22}).

A series of 530 pairs of images acquired with a frequency 2 Hz are
stored for calculating velocity maps and for ensemble and
spatial averaging of turbulence characteristics.
The velocity maps can be considered as independent,
since the maximum correlation time of turbulence is
about 0.15 s, while the time interval between pairs of
images is 0.5 s.

\begin{figure*}[t!]
\centering
\includegraphics[width=8.5cm]{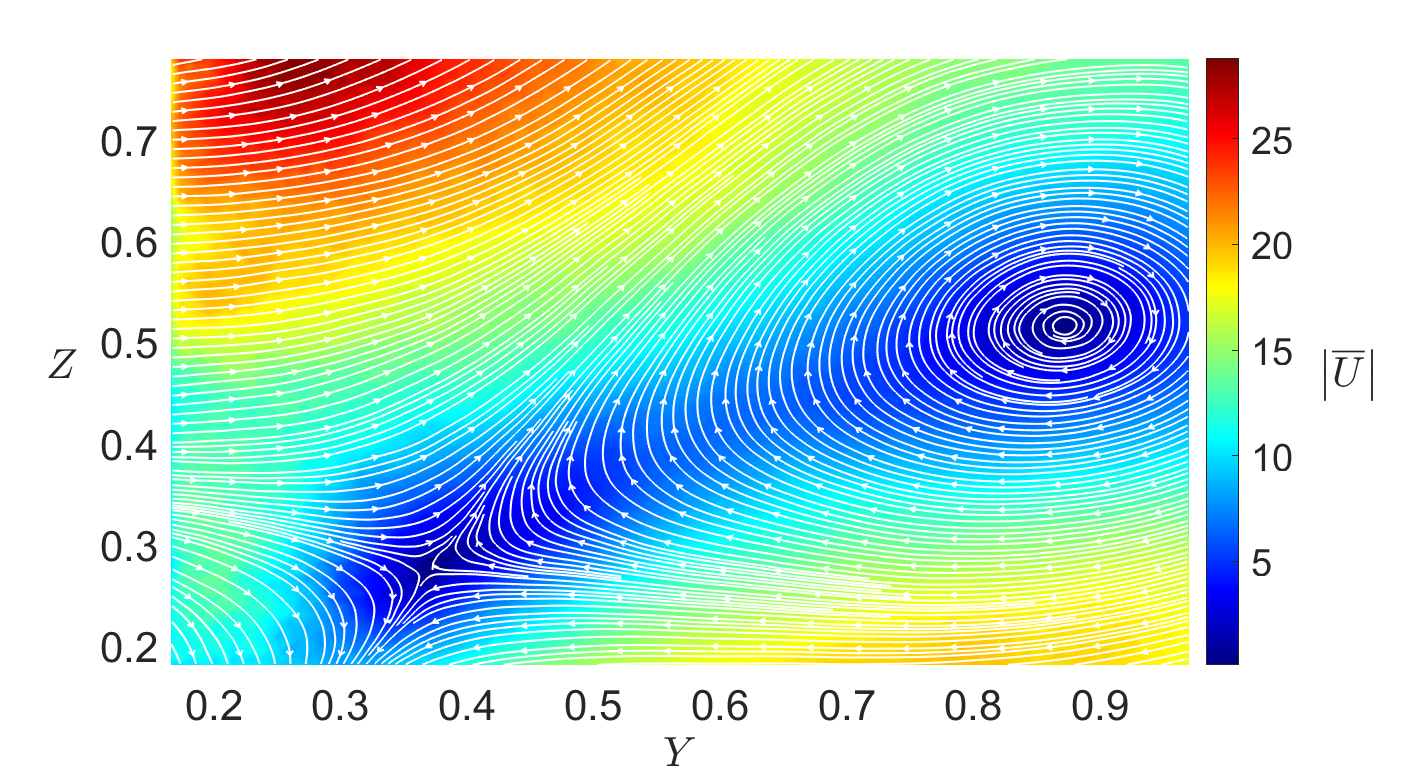}
\includegraphics[width=8.5cm]{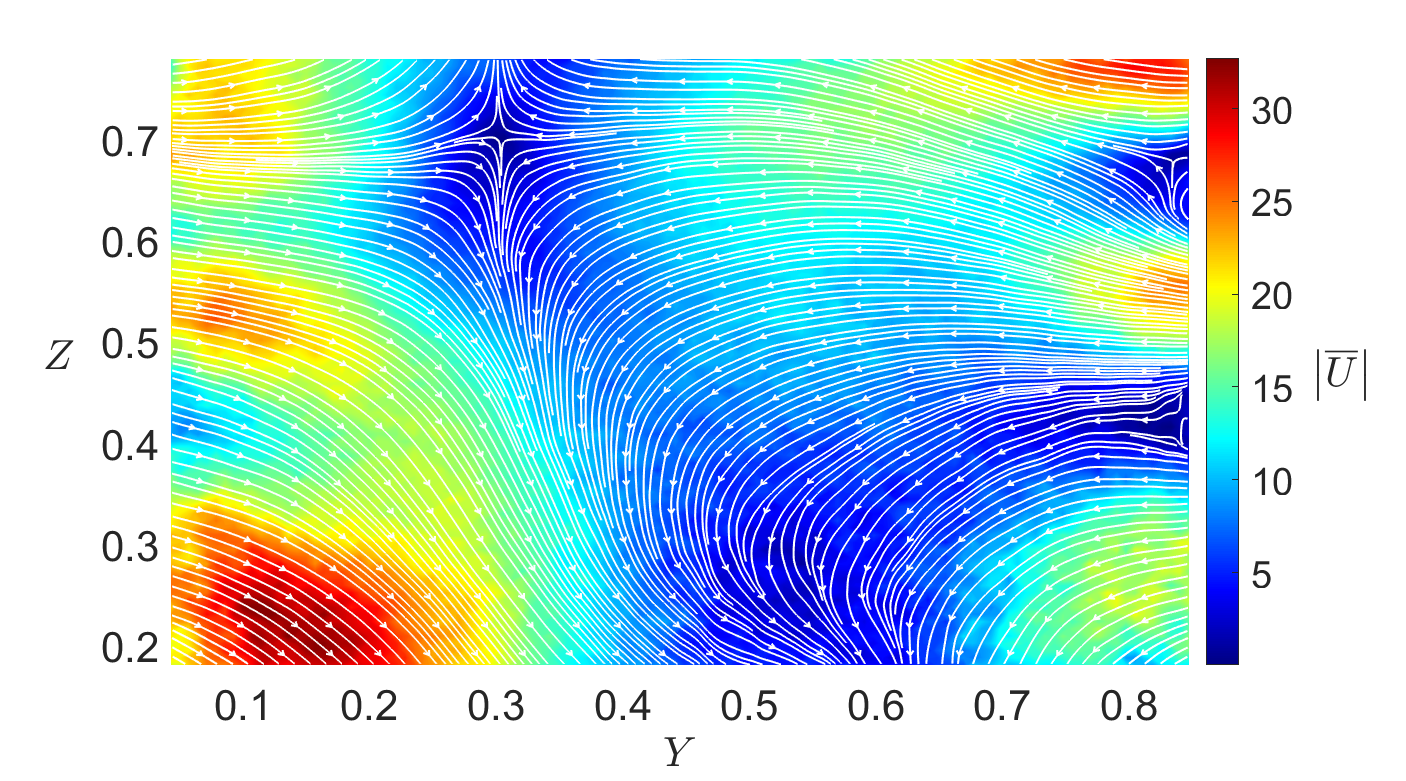}
\caption{\label{Fig2}
Distributions of the mean velocity field $\meanU$
for convective turbulence
forced by one oscillating grid  (left panel) and by two oscillating grids (right panel).
The coordinates $Y$ and $Z$ are normalized by $L_z=26$ cm.
The mean velocity $\meanU$ is measured in cm/s.
}
\end{figure*}

\begin{figure*}[t!]
\centering
\includegraphics[width=8.5cm]{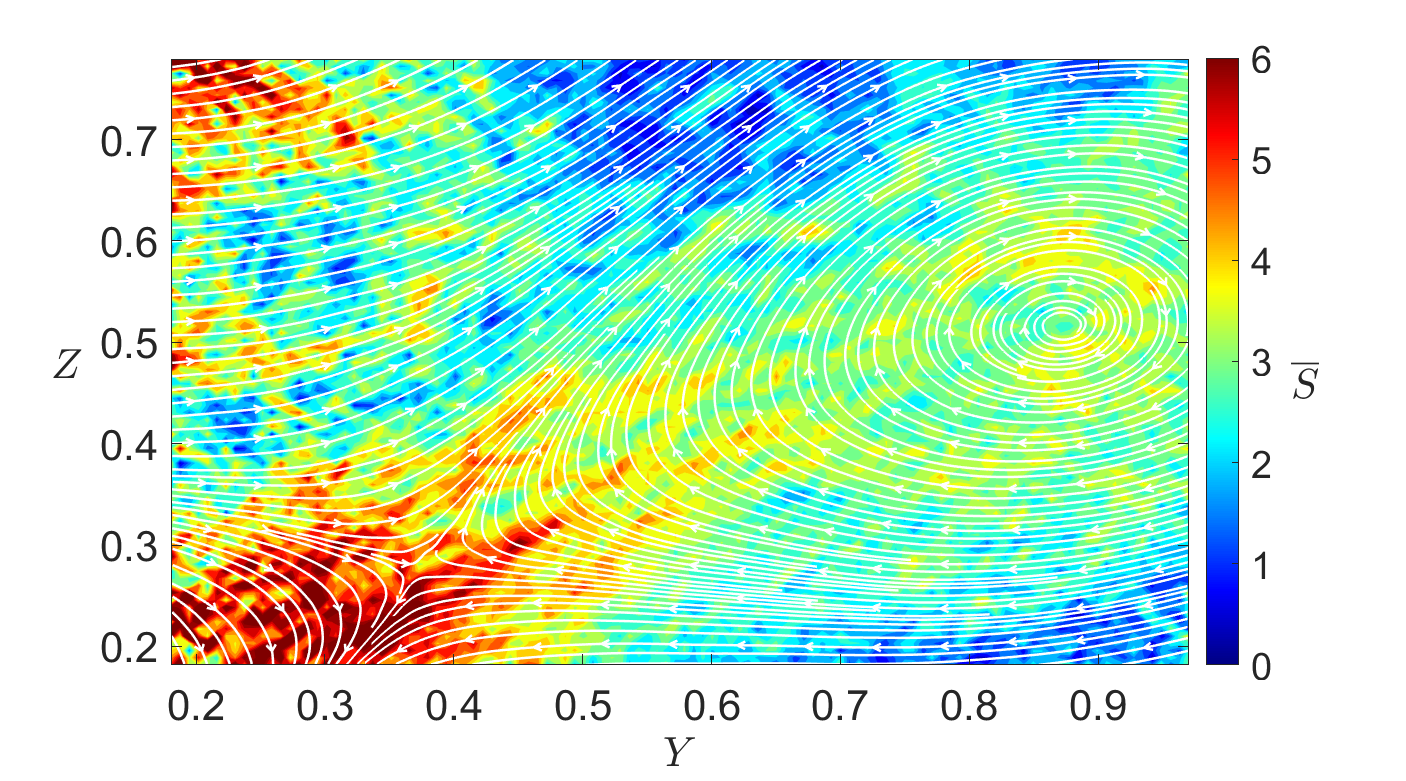}
\includegraphics[width=8.5cm]{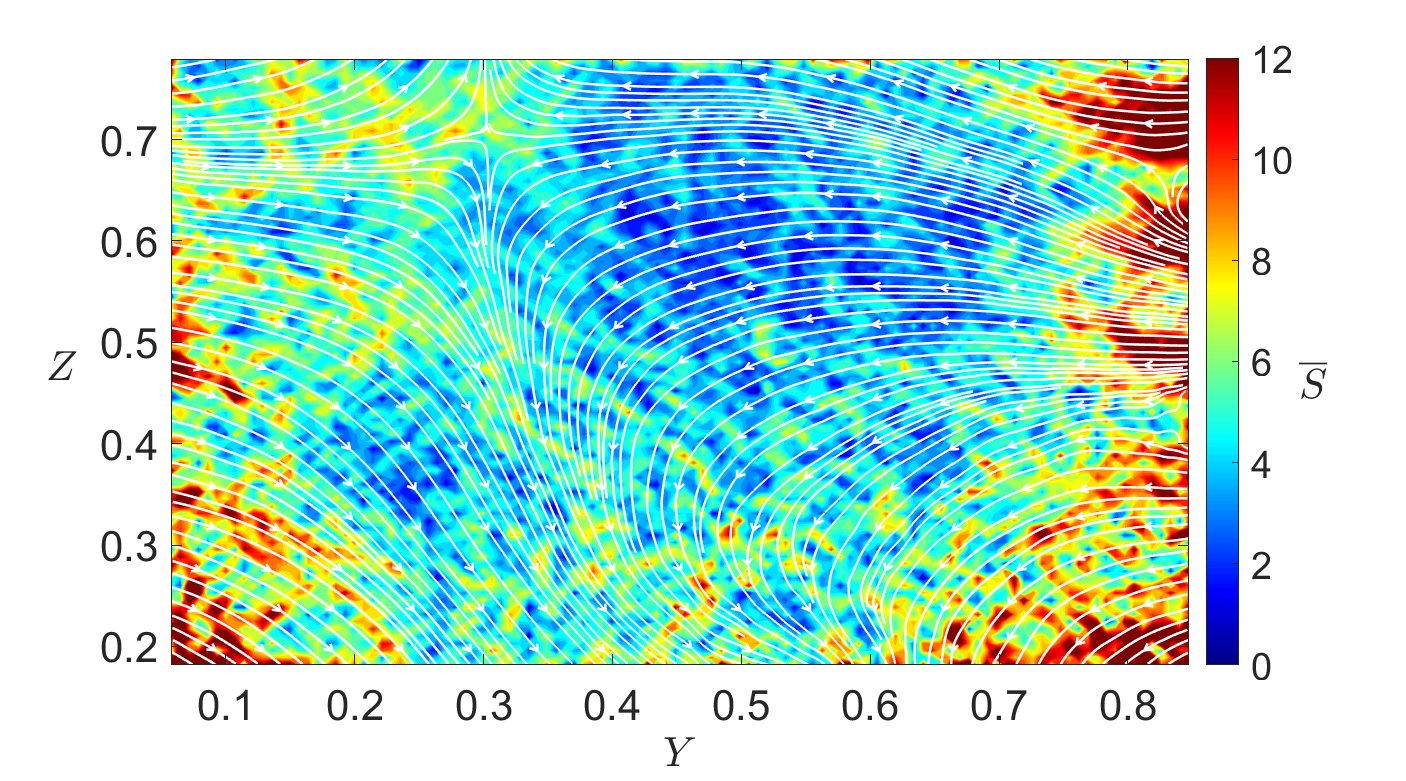}
\caption{\label{Fig3}
Distributions of the mean velocity shear $\meanS=\left[(\nabla_y \meanU_y)^2 + (\nabla_z \meanU_y)^2 + (\nabla_y \meanU_z)^2
+ (\nabla_z \meanU_z)^2\right]^{1/2}$
for convective turbulence
forced by one oscillating grid  (left panel) and by two oscillating grids (right panel).
The streamlines (white) of the mean velocity $\meanUU$ are also superimposed on this distribution.
The coordinates $Y$ and $Z$ are normalized by $L_z=26$ cm,
The mean velocity shear $\meanS$ is measured in s$^{-1}$.
}
\end{figure*}

\begin{figure*}[t!]
\centering
\includegraphics[width=8.5cm]{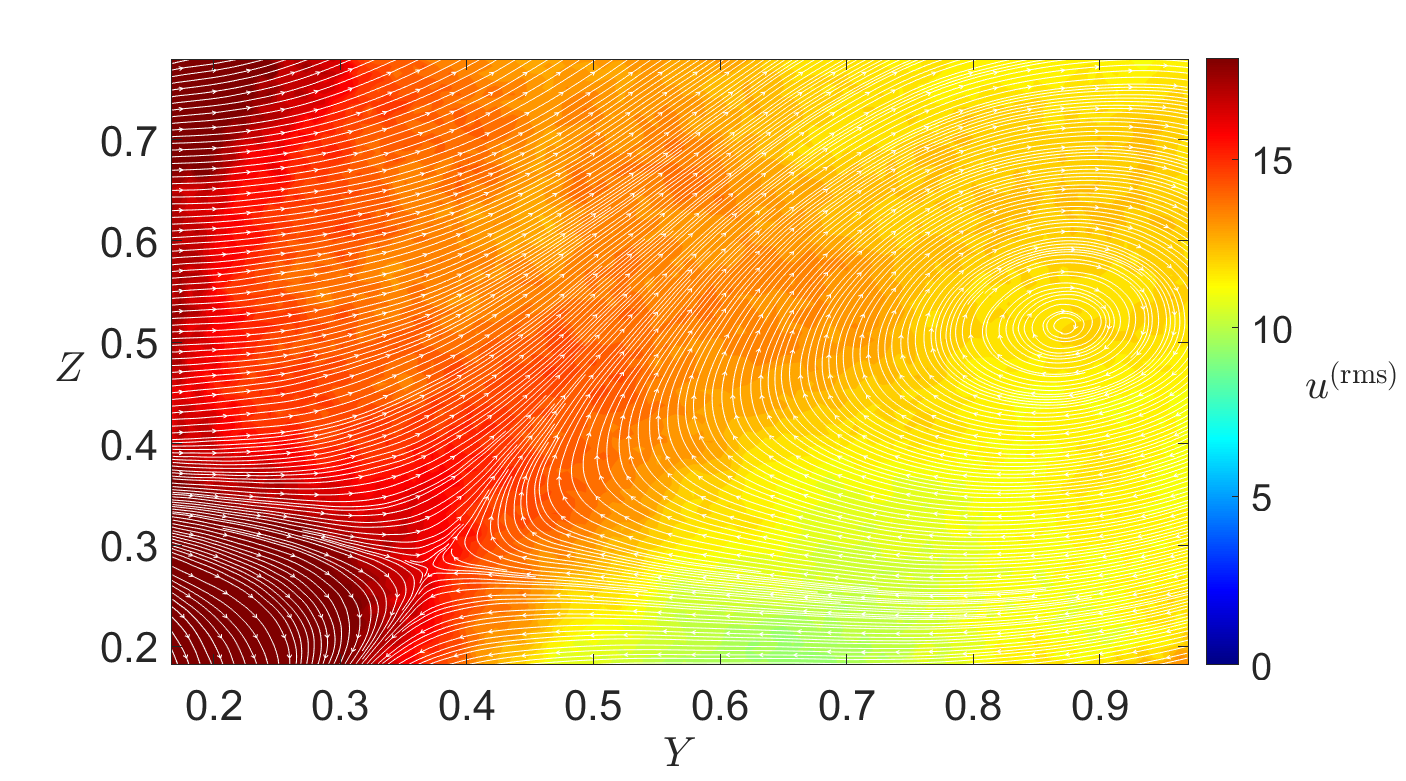}
\includegraphics[width=8.5cm]{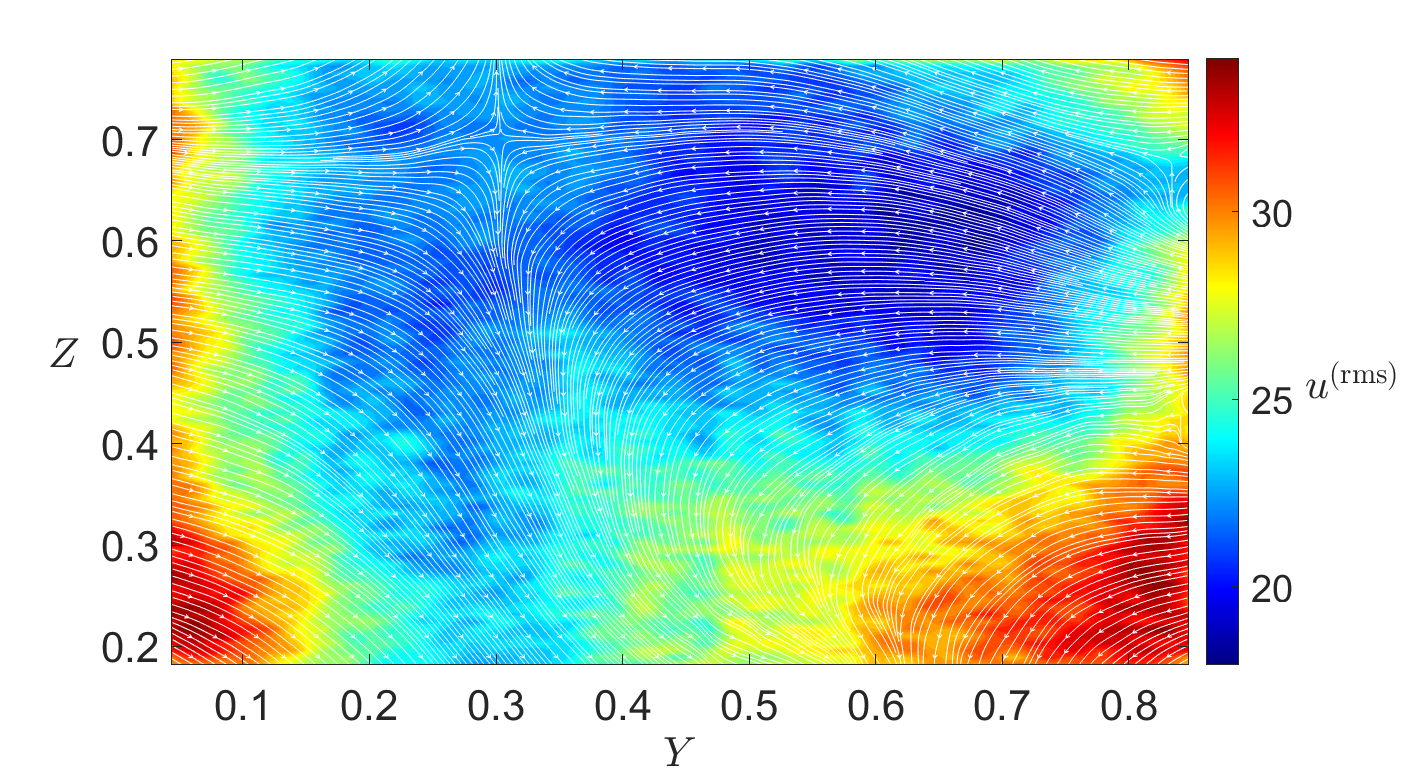}
\caption{\label{Fig4}
Distributions of the turbulent velocity $|u^{\rm (rms)}| = \left[\langle u_y^2 \rangle + \langle u_z^2 \rangle\right]^{1/2}$
for convective turbulence
forced by one oscillating grid  (left panel) and by two oscillating grids (right panel).
The streamlines (white) of the mean velocity $\meanUU$ are also superimposed on this distribution.
The coordinates $Y$ and $Z$ are normalized by $L_z=26$ cm.
}
\end{figure*}

\begin{figure*}[t!]
\centering
\includegraphics[width=8.5cm]{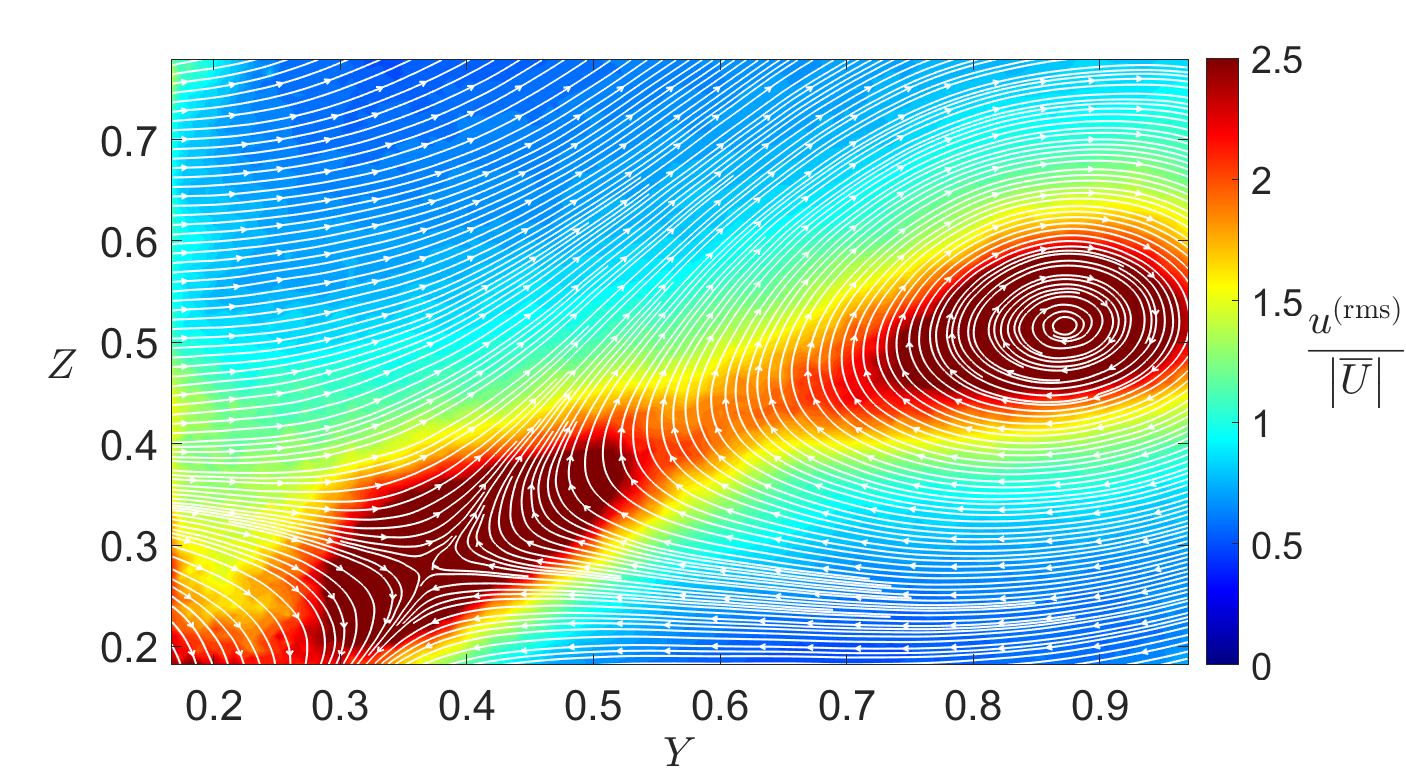}
\includegraphics[width=8.5cm]{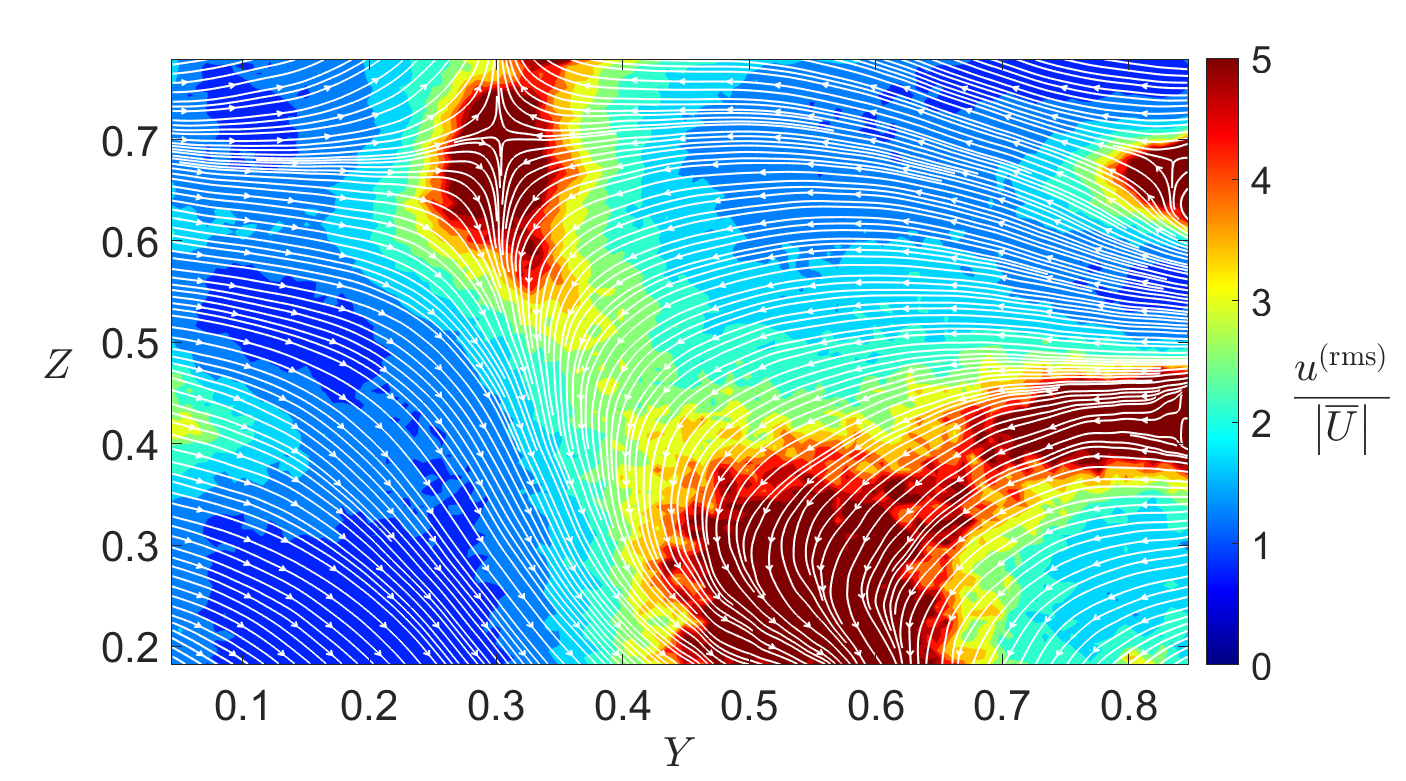}
\caption{\label{Fig5}
Distributions of the ratio $u^{\rm (rms)} / |\meanUU|$
for convective turbulence
forced by one oscillating grid  (left panel) and by two oscillating grids (right panel).
The streamlines (white) of the mean velocity $\meanUU$ are also superimposed on this distribution.
The coordinates $Y$ and $Z$ are normalized by $L_z=26$ cm.
}
\end{figure*}

\begin{figure*}[t!]
\centering
\includegraphics[width=8.5cm]{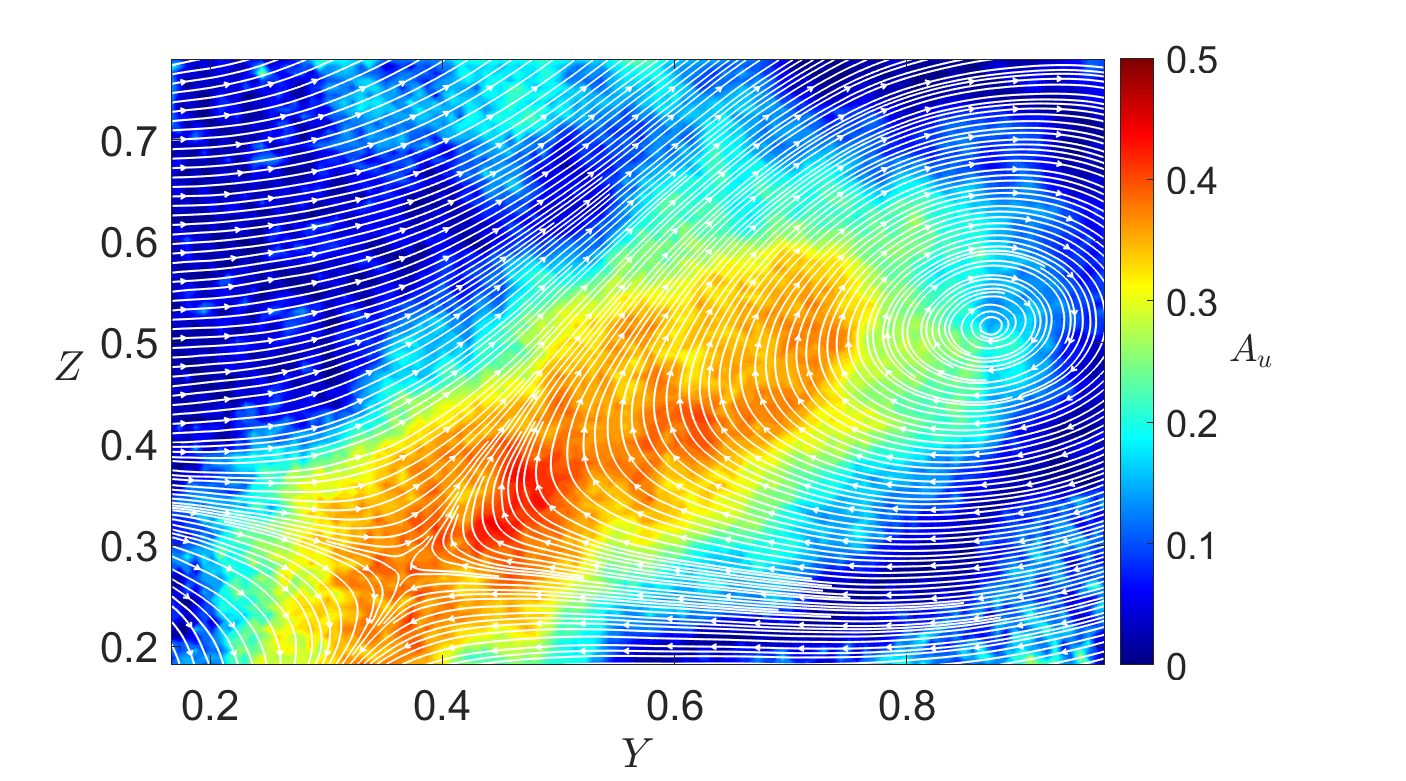}
\includegraphics[width=8.5cm]{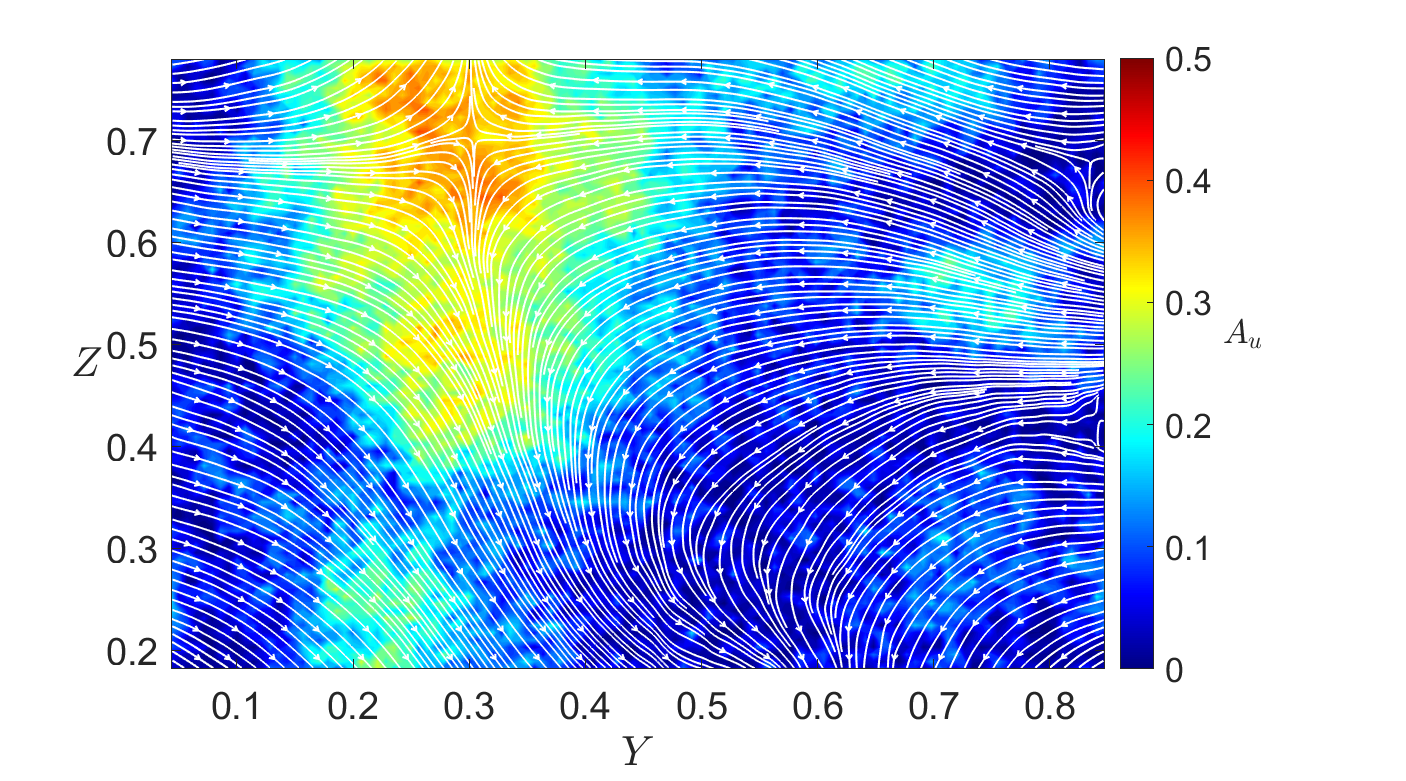}
\caption{\label{Fig6}
Distributions of the anisotropy of turbulent velocity field $A_u =|u_z^{\rm (rms)} / u_y^{\rm (rms)} -1|$
for convective turbulence forced by one oscillating grid  (left panel) and by two oscillating grids (right panel).
The streamlines (white) of the mean velocity $\meanUU$ are also superimposed on this distribution.
The coordinates $Y$ and $Z$ are normalized by $L_z=26$ cm.
}
\end{figure*}

\begin{figure*}[t!]
\centering
\includegraphics[width=8.5cm]{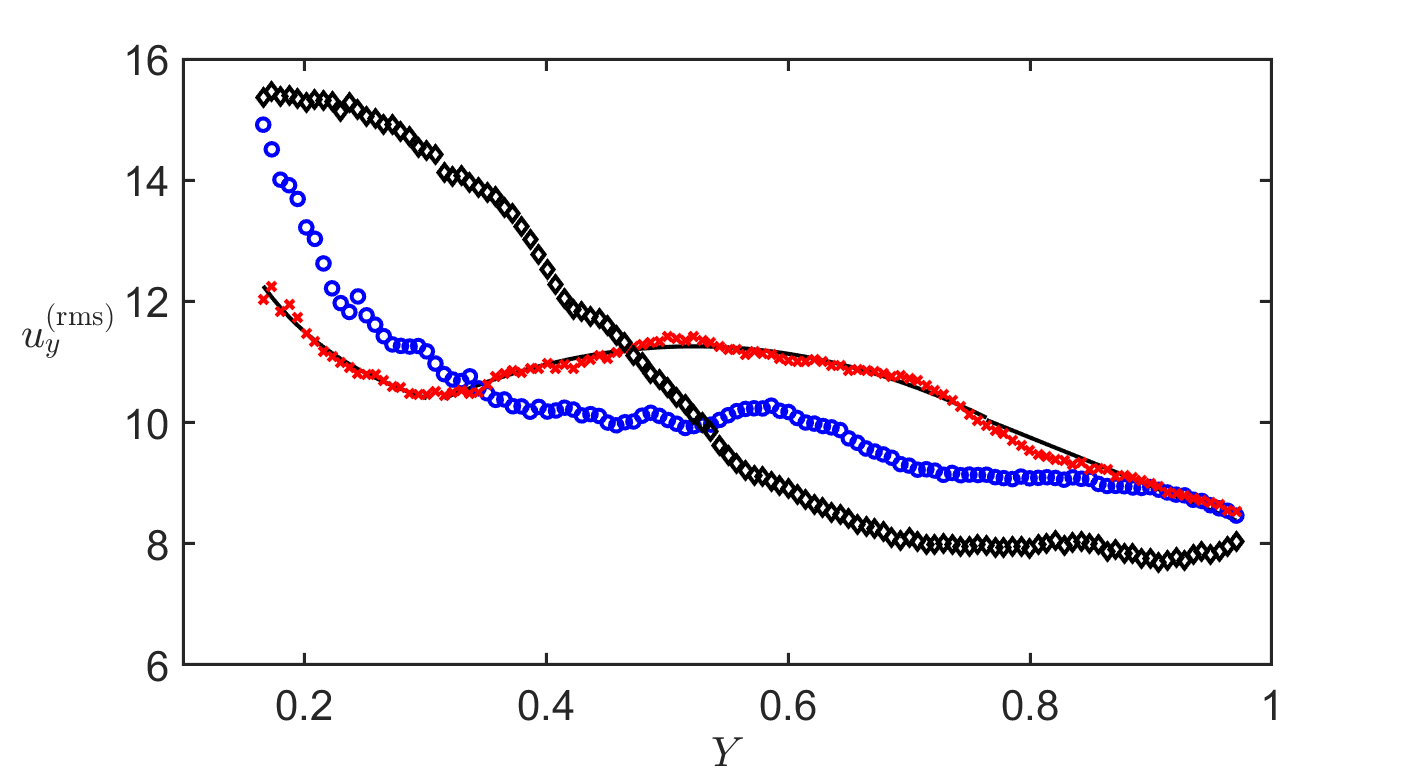}
\includegraphics[width=8.5cm]{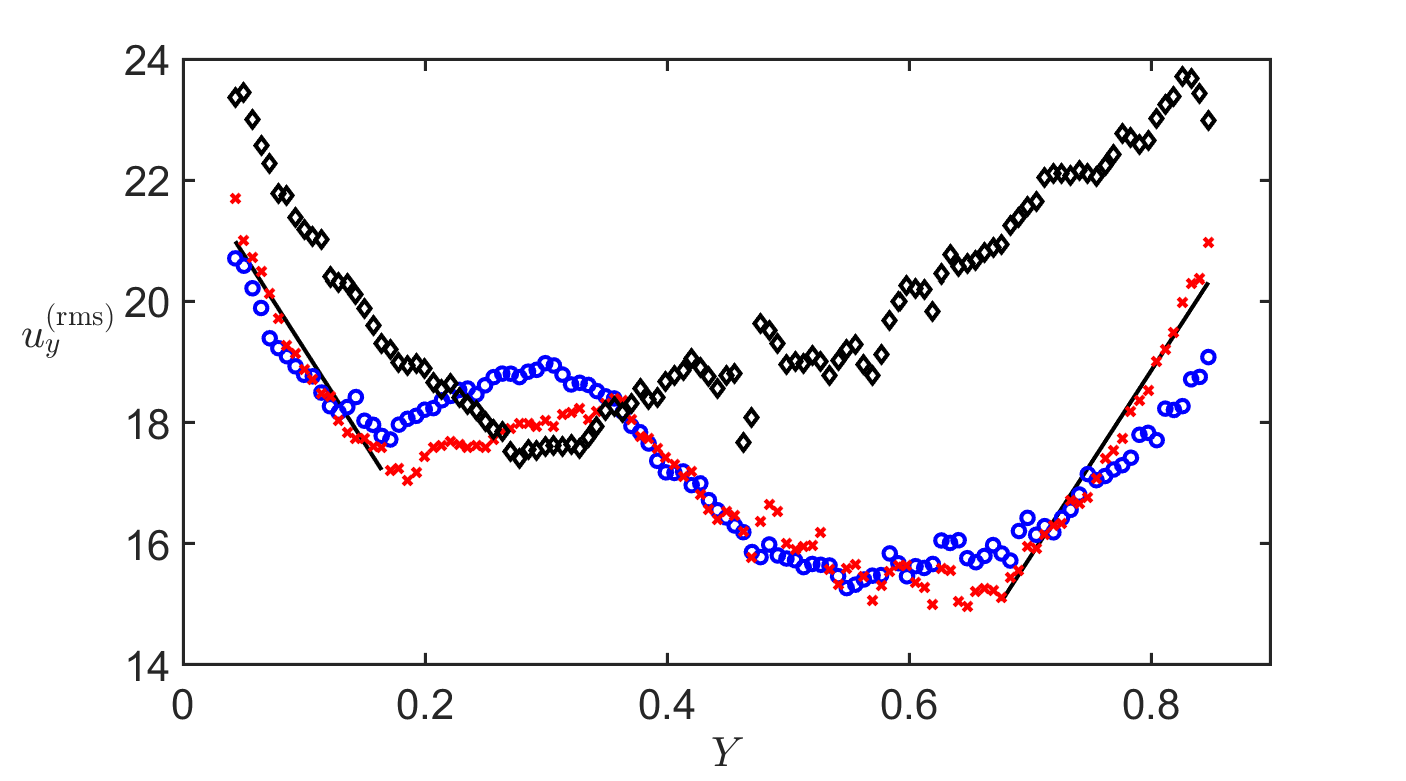}
\caption{\label{Fig7}
Dependencies of  the turbulent velocity $u_y^{\rm (rms)}(Y)$
on the horizontal coordinate $Y$ in the core flow averaged over
various ranges $Z$ for convective turbulence
forced by one oscillating grid  (left panel) and by two oscillating grids (right panel).
Fitting curves are shown by the solid lines.
The averaging is over $Z = $ 4.7-9.2 cm (blue, circles); 9.2-14.7 cm
(red, crosses); 14.7-20.3 cm (black, diamonds).
The velocity is measured in cm/s and the coordinates $Y$ is normalized by $L_z=26$ cm.
}
\end{figure*}

\begin{figure*}[t!]
\centering
\includegraphics[width=8.5cm]{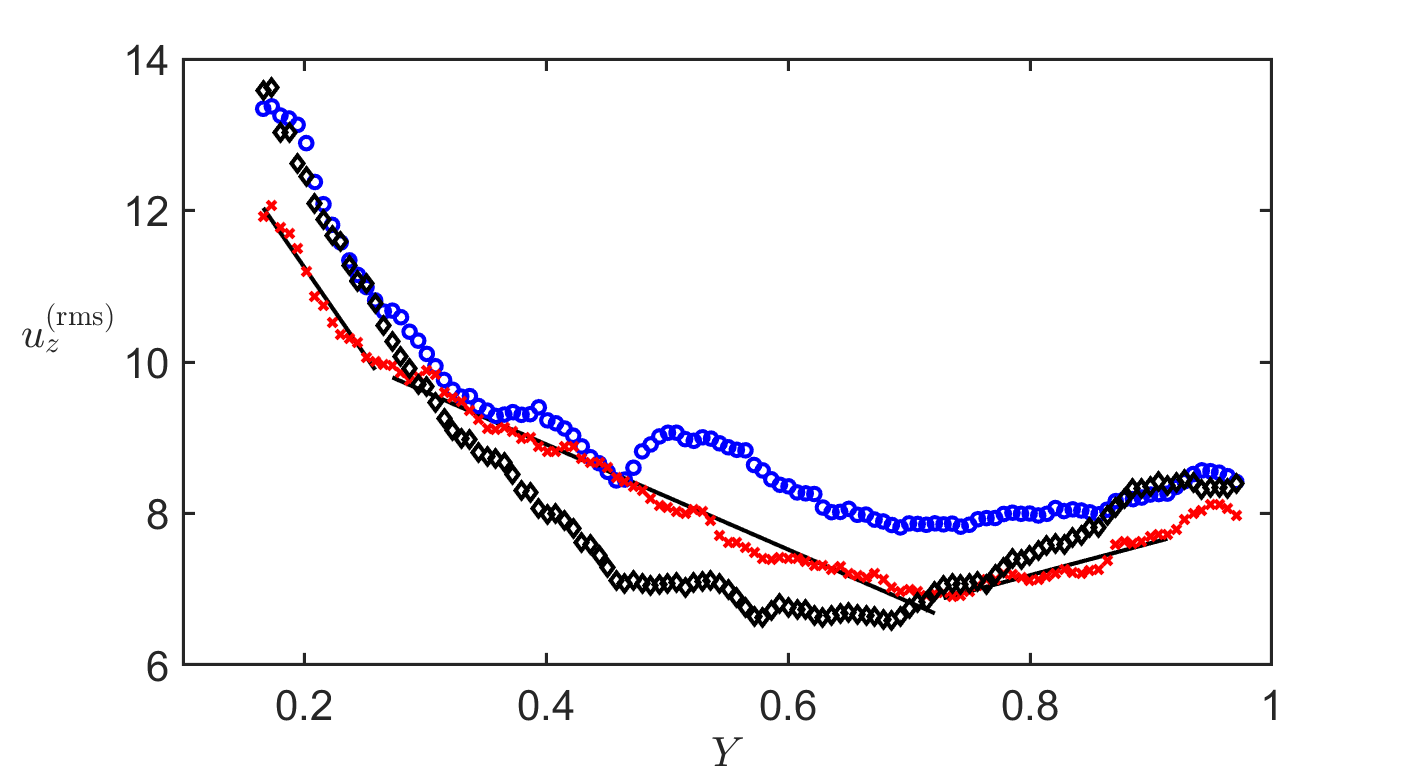}
\includegraphics[width=8.5cm]{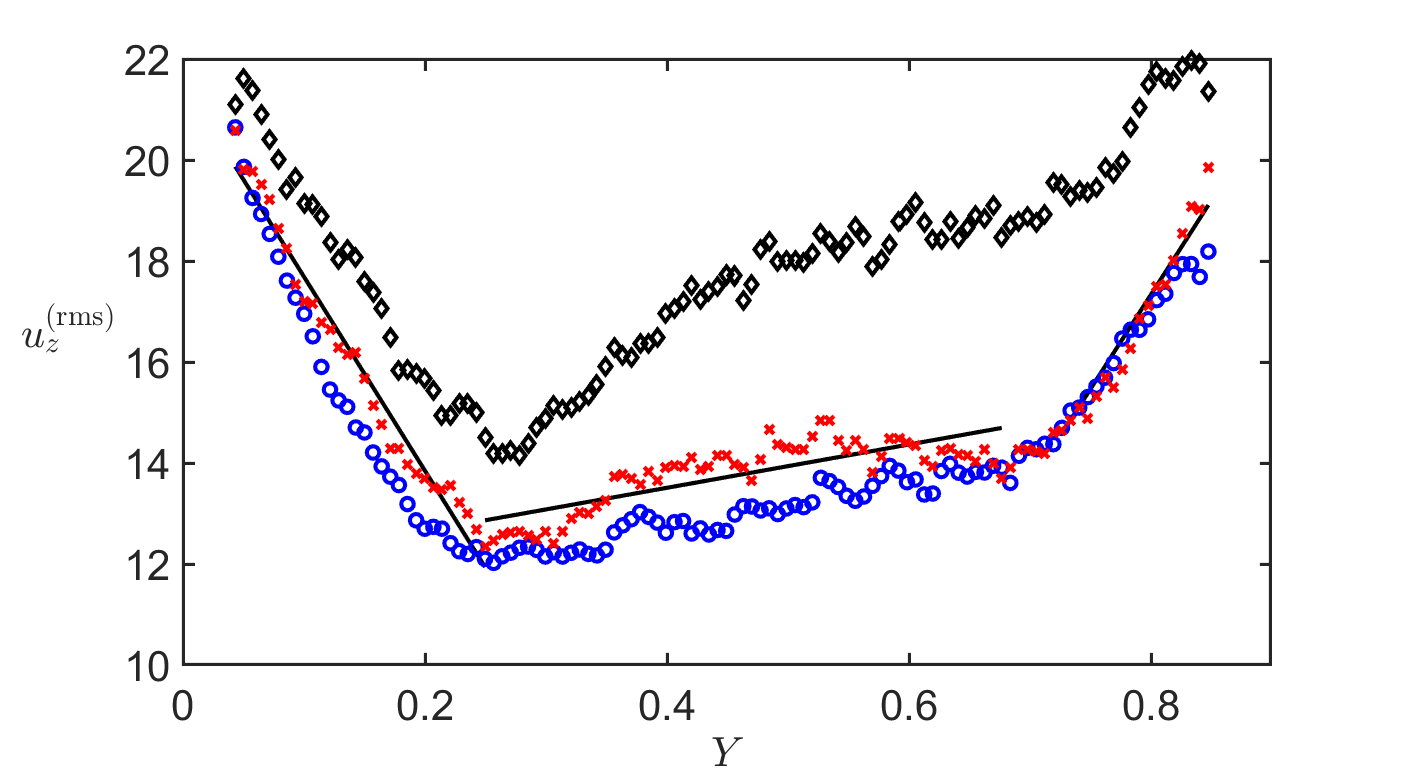}
\caption{\label{Fig8}
Dependencies of  the turbulent velocity $u_z^{\rm (rms)}(Y)$
on the horizontal coordinate $Y$ in the core flow averaged over
various ranges $Z$ for convective turbulence
forced by one oscillating grid  (left panel) and by two oscillating grids (right panel).
Fitting curves are shown by the solid lines.
The left panel: the averaging is over $Z = $ 5.2-9.3 cm (blue, circles); 9.2-14.3 cm
(red, crosses); 14.3-18.6 cm (black, diamonds).
The right panel: the averaging is over $Z = $ 5.3-9.4 cm (blue, circles); 9.4-14.2 cm
(red, crosses); 14.2-18.4 cm (black, diamonds).
The velocity is measured in cm/s and the coordinates $Y$ is normalized by $L_z=26$ cm.
}
\end{figure*}

\begin{figure*}[t!]
\centering
\includegraphics[width=8.5cm]{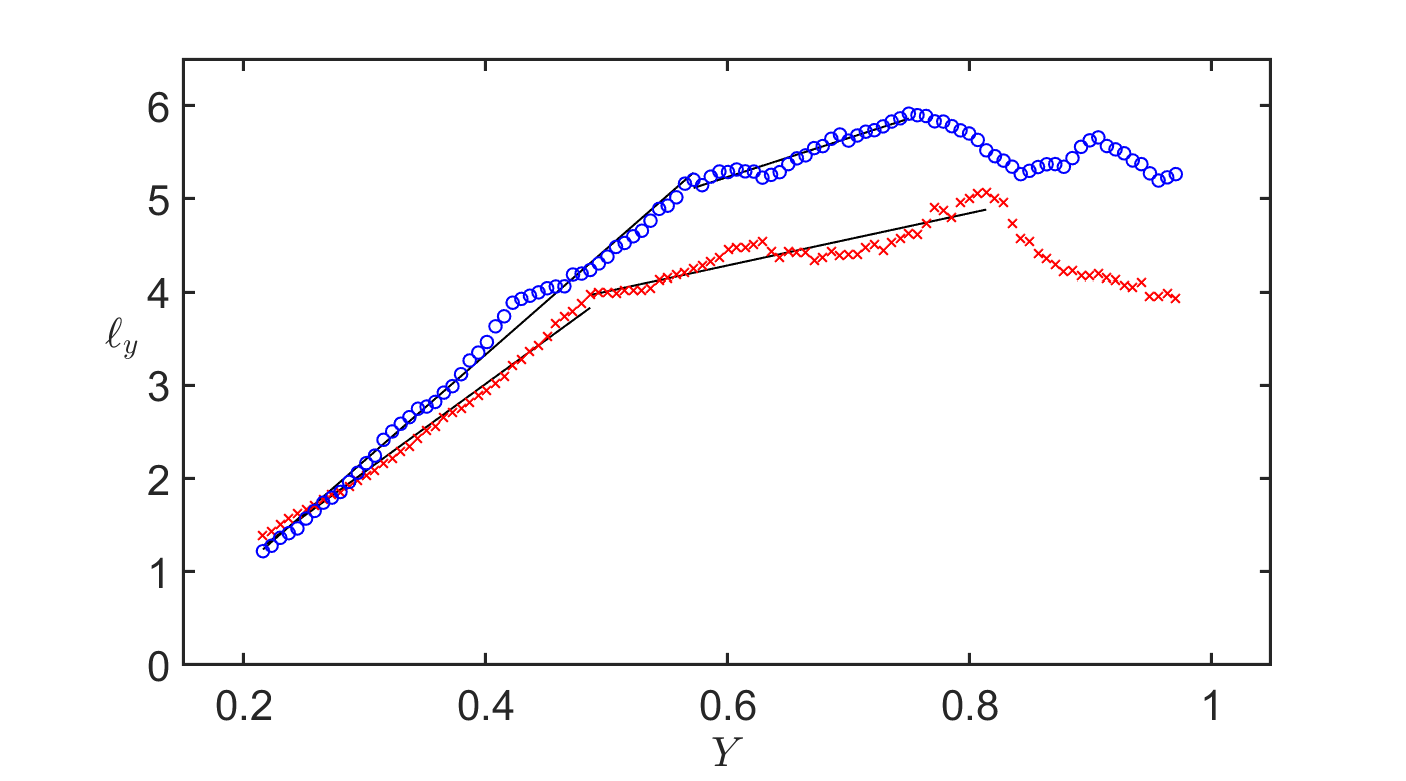}
\includegraphics[width=8.5cm]{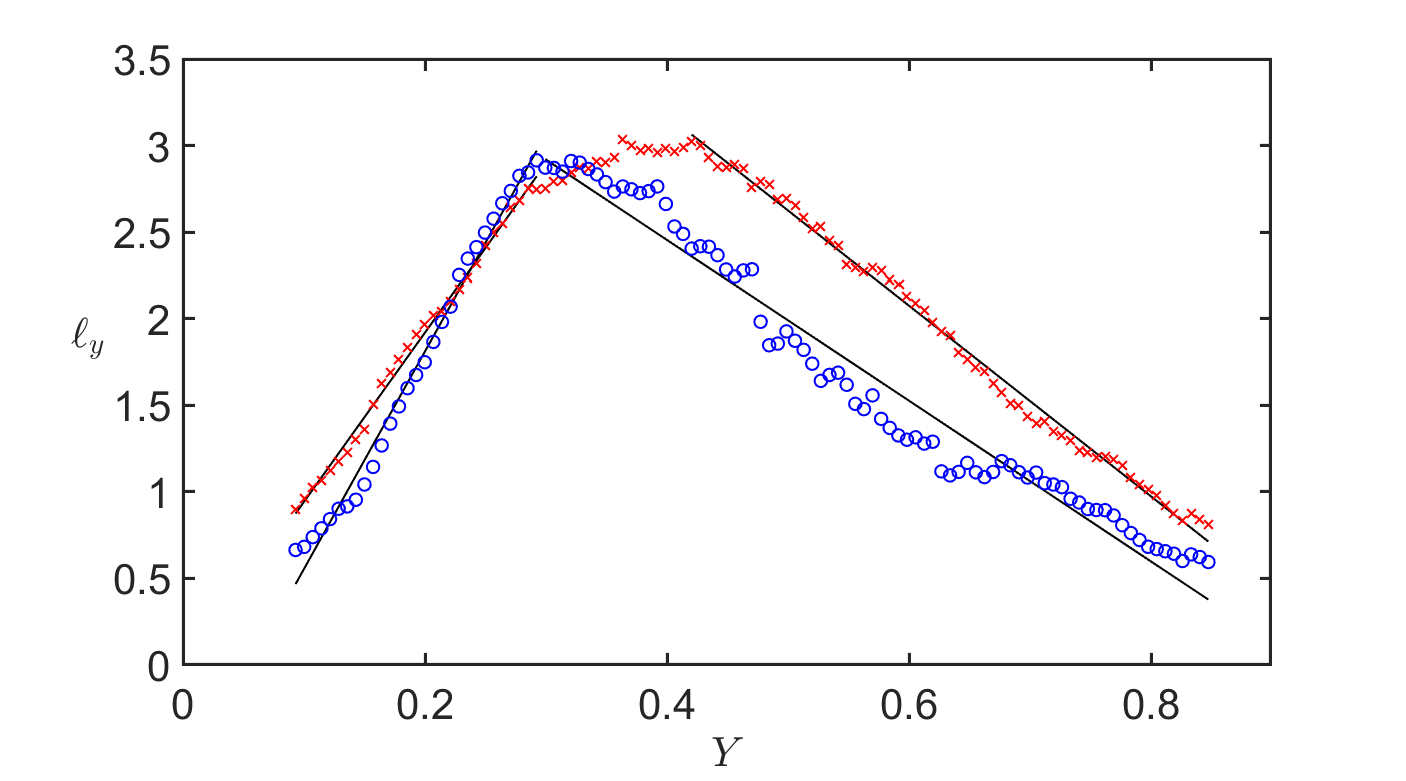}
\caption{\label{Fig9}
Dependencies of  the horizontal integral turbulence scale $\ell_y(Y)$
on the horizontal coordinate $Y$ in the core flow averaged over
$Z$ for isothermal (red) and convective (blue) turbulence
forced by one oscillating grid  (left panel) and by two oscillating grids (right panel).
Fitting curves are shown by the solid lines.
The integral turbulence scale is measured in cm and the coordinate $Y$ is normalized by $L_z=26$ cm.
}
\end{figure*}

\begin{figure*}[t!]
\centering
\includegraphics[width=8.5cm]{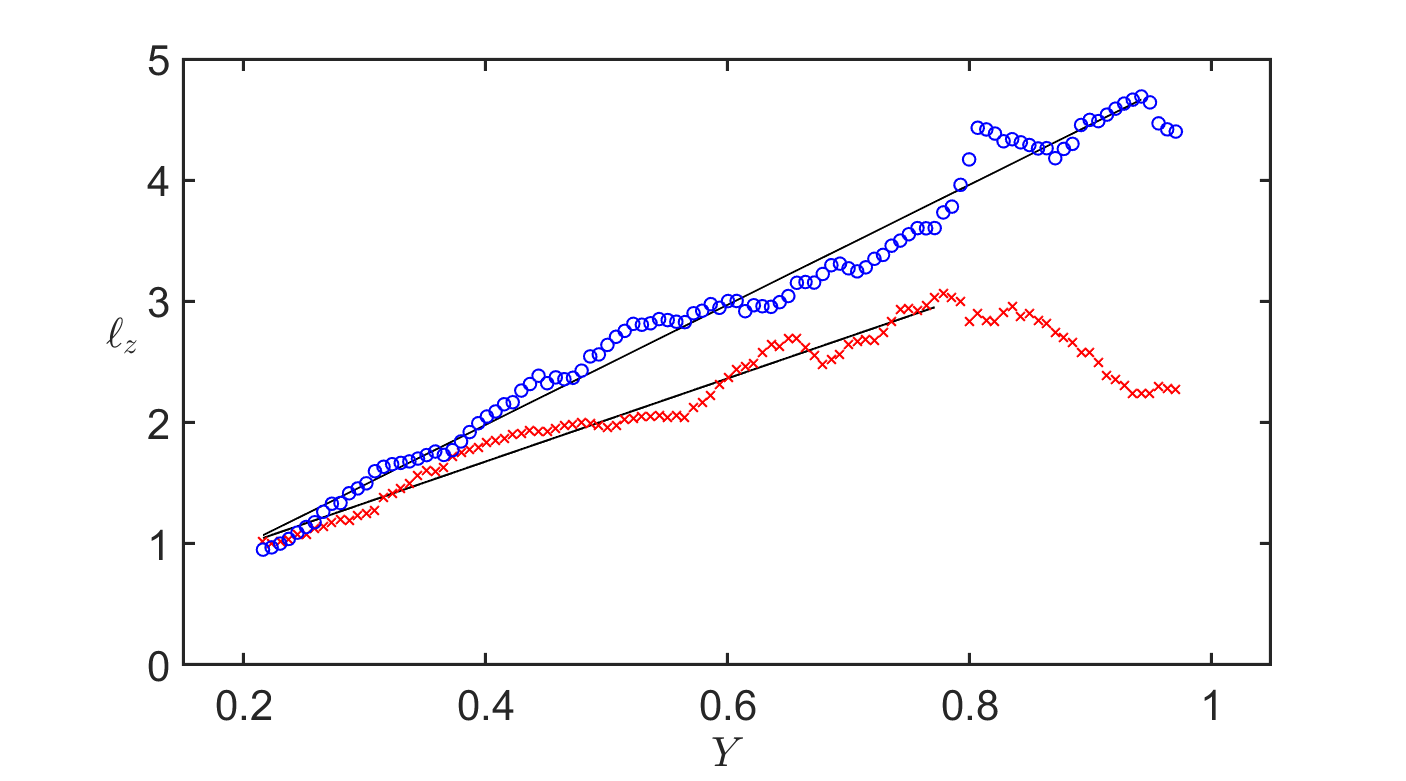}
\includegraphics[width=8.5cm]{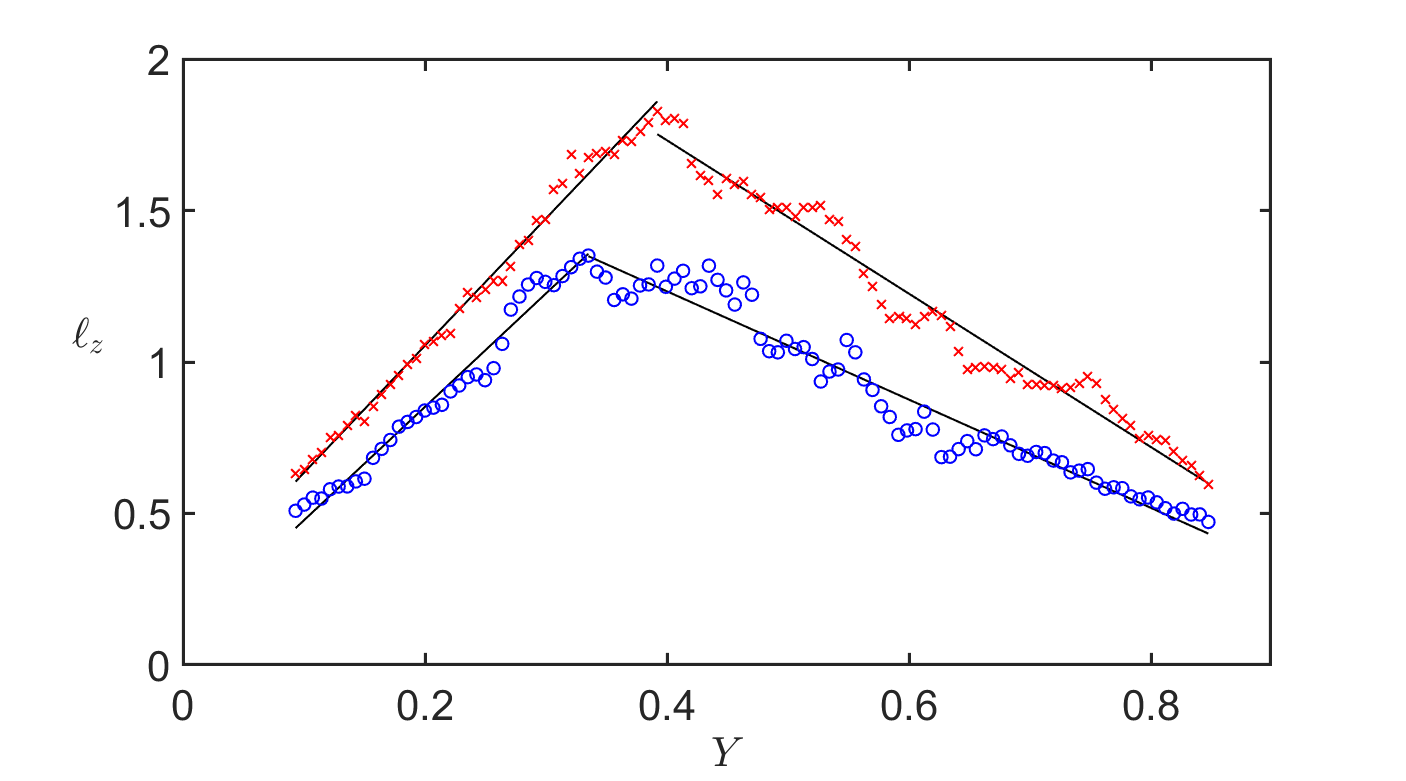}
\caption{\label{Fig10}
Dependencies of  the vertical integral turbulence scale $\ell_z(Y)$
on the horizontal coordinate $Y$ in the core flow averaged over
$Z$ for isothermal (red) and convective (blue) turbulence
forced by one oscillating grid  (left panel) and by two oscillating grids (right panel).
Fitting curves are shown by the solid lines.
The integral turbulence scale is measured in cm and the coordinates $Y$ is normalized by $L_z=26$ cm.
}
\end{figure*}

The velocity fields in the experiments have been measured  in a flow
domain $209.1 \times 155.4$ mm$^2$ with a spatial
resolution of $1376 \times 1024$ pixels, so that
a spatial resolution 151 $\mu$m /pixel have been achieved.
We analyse the velocity field in the probed region
with interrogation windows of $16
\times 16$ pixels.
Using the velocity measurements,
various mean-field  (see Figs.~\ref{Fig2}--\ref{Fig3}) and turbulent (see Figs.~\ref{Fig4}--\ref{Fig11}) characteristics
have been obtained in our experiments.
In particular, we determine the mean velocity field and mean velocity shear,
the root mean square (r.m.s.) velocities, the turbulent anisotropy parameter,
the two-point correlation functions of velocity field
and the integral scales of turbulence.

The integral length scales of turbulence $\ell_y$ and $\ell_z$ are calculated
in the horizontal $Y$ and the vertical $Z$ directions from the
two-point correlation functions of the fluid velocity fluctuations.
In particular, to determine the integral turbulent length scales
we use two approaches which yield the similar results.
In the first approach, we integrate the normalized two-point correlation function of the fluid velocity fluctuations
over the distance between two points which varies from zero to the point, where the
correlation function vanishes.
In the second method, the integral turbulent length scale is defined as the characteristic distance
at which the normalized two-point correlation function $\propto \exp(-r^2/\ell_0^2)$ decreases  in $e$ times.

The mean and r.m.s. velocities are determined for
every point of a velocity map by averaging over 530 independent maps.
In the experiments we evaluated the
variability between the first and the last 20
velocity maps of the series of the measured
velocity field. Since very small variability is
found, these tests show that 530 velocity maps
contain enough data to obtain reliable
statistical estimates.
The size of the probed region does not affect our results.
Increasing the number of  velocity maps
we obtain the same results.

In our experiments, the spatial distributions for particles having the diameters 0.7 $\mu m$ and 10 $\mu m$
are obtained by the PIV system
using the effect of the Mie light scattering by particles \cite{guib01}.
In particular,
the mean intensity of scattered light is determined
in $85 \times 64$ interrogation windows with the size
$32 \times 32$ pixels (in convective turbulence
forced by one oscillating grid),
and in $65 \times 51$ interrogation windows with the size
$32 \times 32$ pixels (in convective turbulence
forced by two oscillating grids),
with 50 \% overlap (for better averaging data and smoothness).
This allows us to find the vertical distribution of the intensity of
the scattered light in 80 vertical strips composed of
64 interrogation windows.
Here we take into account that
the light radiation energy flux scattered
by small particles is given by $E_s \propto E_0 \Psi(\pi d_{\rm p}/\lambda; a_0;n)$, where
$\Psi$ is the scattering function, $d_{\rm p}$ is the particle diameter, $\lambda$ is the
wavelength, and $a_0$ is the index of refraction.
The energy flux incident at the particle is given by
$E_0 \propto \pi d_{\rm p}^2 / 4$.
For $\lambda > \pi d_{\rm p}$, the  scattering function $\Psi$ is determined
by the Rayleigh's law, $\Psi \propto d_{\rm p}^4$.
For $\lambda < \pi d_{\rm p}$, the scattering function $\Psi$ is independent of
the particle diameter and the wavelength.
In a general case, the scattering function $\Psi$ is determined by the Mie
equations \cite{BH83}.
The light radiation energy flux scattered
by small particles is given by $ E_s \propto E_0 \, n \,
(\pi d_{\rm p}^2 / 4) $, so
that the scattered light energy flux
incident on the charge-coupled device (CCD) camera probe is proportional
to the particle number density $n$.

The scattered radiation intensity recorded by the CCD camera
at a given spatial location is proportional to the local particle number density.
Therefore, intensity ratios obtained at that location in different measurements
directly represent the corresponding ratios of particle number densities.
For the normalization of the scattered light intensity $E^T$ obtained in
a temperature-stratified turbulence,
the distribution of the scattered light intensity $E$ measured in the
isothermal case is used under the same conditions.
The distribution of the scattered light intensity averaged over
a vertical coordinate is independent of the particle number
density in the isothermal flow.
Therefore, using this normalization, we can characterize the spatial distribution of particle
number density $n \propto E^T /E$ in the non-isothermal turbulence.
A series of 530 images acquired with a frequency 4 Hz are
stored for ensemble and spatial averaging of particle number density distributions.

For experimental study of turbulent thermal diffusion of inertial particles, we
use  borosilicate hollow glass particles having an approximately
spherical shape, a mean diameter of $10 \, \mu$m
and the material density $ \rho_p \approx 1.4 $ g/cm$^3$.
These particles have been injected in the chamber using an air jet
to improve particle mixing and prevent from particle agglomeration.

We use a custom-made acoustic feeding device for injecting particles
into the flow comprising an acrylic glass chamber with a size of 9 $\times$ 9 $\times$ 4 cm$^3$.
Two plastic slabs inside the chamber are used as air guides to achieve optimal flow
with entrained particles.
Particle dispensation zone (a disk of 25 mm diameter and 5 mm thickness)
is located at the bottom of the chamber.
A standard woofer (oval 2 $\times$ 3.5") at a frequency of 220 Hz sways a latex membrane
on which particles are loaded.
A cylindrical cavity is used to contain the particles on the latex membrane.
The batch of particles on the membrane should roughly fill the cavity.
When the membrane vibrates, particles are entrained into air.
Particle feeding device has a pressurized air inlet with bellows having a 8 mm diameter tube
with standard quick release connector.
The entrained particles leave the chamber with a stream of air through the outlet.

Submicron particles (0.7 $\mu$m) are used as tracer particles for the PIV measurements of the velocity field.
In separate experiments, submicron particles (0.7 $\mu$m) are used to study turbulent thermal diffusion
of noninertial particles.
On the other hand, larger particles (10 $\mu$m) are used in separate experiments
to study turbulent thermal diffusion of inertial particles.
Particles do not affect the fluid flow because the mass-loading parameter is small ($m_{\rm p} n \ll \rho$).
The particles are injected in the chamber in different locations and directions
(2 inlets are on the upper wall and 2 inlets are on the bottom wall of the chamber), to spread the particles uniformly
in the flow domain.

The measurement technique and data processing procedure described in this section
are similar to those used by us  in various experiments with
turbulent convection \cite{BEKR09,EEKR11,BELR20,SKRL22,EKRL23}
and stably stratified turbulence \cite{EEKR13,CEKR14,EKRL22},
in the experiments to study turbulent thermal
diffusion of noninertial particles in a homogeneous turbulence \cite{BEE04,EEKR04,EEKR06a,AEKR17},
mixing of particles in inhomogeneous turbulence \cite{EHSR09,EKRL22,EKRL23},
as well as for investigation of small-scale particle clustering \cite{EKR10}.

The reliability of the experimental measurements has been assessed by considering the main sources of uncertainty associated with the velocity, temperature, and particle number density fields.
The velocity field was obtained using the PIV system, and its quality has been evaluated using the standard $Q$ factor criterion, defined as the ratio between the primary and secondary peaks of the cross-correlation function \cite{AD91,RWK07,W00}. The spatially averaged $Q$ factor was found to be approximately 1.6 or higher across the dataset, indicating reliable cross-correlation and sufficient signal-to-noise ratio for accurate velocity estimation. In addition, standard validation procedures have been applied to remove spurious vectors and ensure the robustness of the velocity field. Since key quantities such as the Reynolds number and integral length scales are derived from the velocity field, the demonstrated quality of the PIV data supports the reliability of all subsequent analyses.

The temperature field has been measured using an array of thermocouples. The uncertainty in the temperature measurements is estimated to be of the order of $ \pm 0.1$--$0.3$ K, based on sensor accuracy and measurement conditions. As the analysis is based on time-averaged temperature fields, random noise is significantly reduced. While small uncertainties may arise from spatial positioning and interpolation, particularly in regions with strong gradients, these effects are minor compared to the overall temperature variations across the domain and do not affect the observed trends.

The particle number density is determined from the ratio of the mean scattered light intensities in the non-isothermal and isothermal cases, based on frame-averaged CCD images. The total uncertainty in the particle number density is evaluated using first-order error propagation (the $\delta$ method), accounting for the independent contributions of both measurements \cite{BR03}.
The resulting characteristic uncertainty in the particle number density measurements varies from $4 \%$
to $8 \%$ (95 $\%$ confidence interval), indicating a satisfactory level of experimental accuracy.
Overall, the combined uncertainties in the velocity, temperature, and particle number density measurements
are small relative to the observed variations in the flow and do not affect the main conclusions of the study.

\section{Experimental  results}
\label{sect4}

The main goal of this paper is to investigate the phenomenon of turbulent thermal diffusion of inertial particles.
In this section, we discuss the experimental results on formation of large-scale clusters of inertial particles
in a small-scale convective turbulence forced by one or two oscillating grids in the airflow
(see Fig.~\ref{Fig1}).
To demonstrate the existence of the phenomenon of turbulent thermal diffusion,
it is important to have turbulence with large Reynolds numbers.
So below we present the results on fluid turbulence only for illustration without deep
quantitative analysis, but focusing more on turbulent transport of inertial particles.

In Fig.~\ref{Fig2} we show the mean velocity field in the core flow
for convective turbulence forced by one oscillating grid  (left panel) and by two oscillating grids (right panel).
The oscillating grids do not completely destroy the large-scale circulations,
but their structures are strongly deformed.
The mean velocity patterns in these experiments contain
one or two large-scale circulations.
The large-scale velocity shear inside the large-scale circulation is not small,
see Fig.~\ref{Fig3} with the distributions of the mean velocity shear
$\meanS=\left[(\nabla_y \meanU_y)^2 + (\nabla_z \meanU_y)^2 + (\nabla_y \meanU_z)^2
+ (\nabla_z \meanU_z)^2\right]^{1/2}$ for convective turbulence.
In particular, the shear parameter in the regions with the maximum large-scale velocity shear
in the experiments with convective turbulence forced by one oscillating grid
is about $\meanS^{\,({\rm max})} \tau_0 \approx 0.75$ (where $\tau_0 \approx 0.12$ s),
while in the experiments with convective turbulence forced by two oscillating grids,
it is about $\meanS^{\,({\rm max})} \tau_0 \sim 0.4$ (where $\tau_0 \approx 0.033$ s).

\begin{figure*}[t!]
\centering
\includegraphics[width=8.5cm]{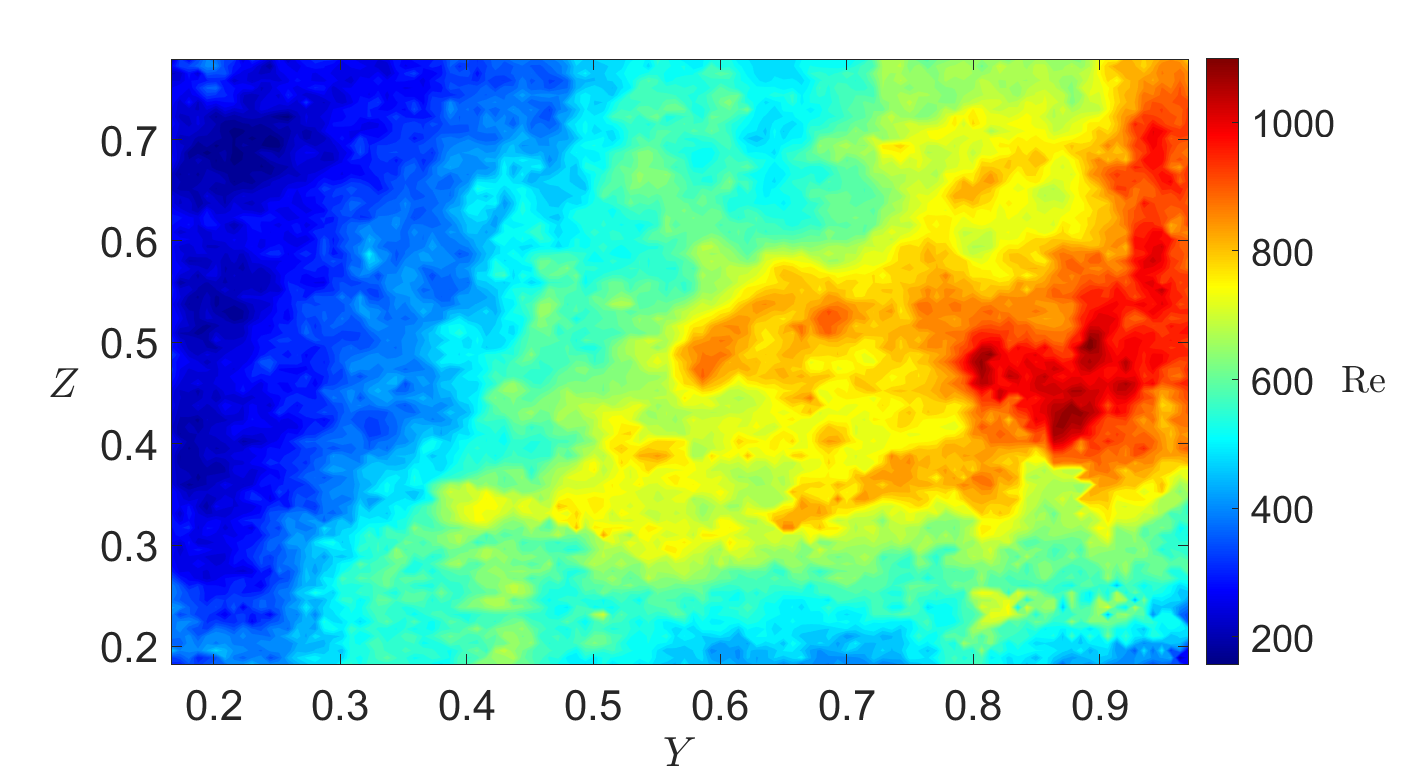}
\includegraphics[width=8.5cm]{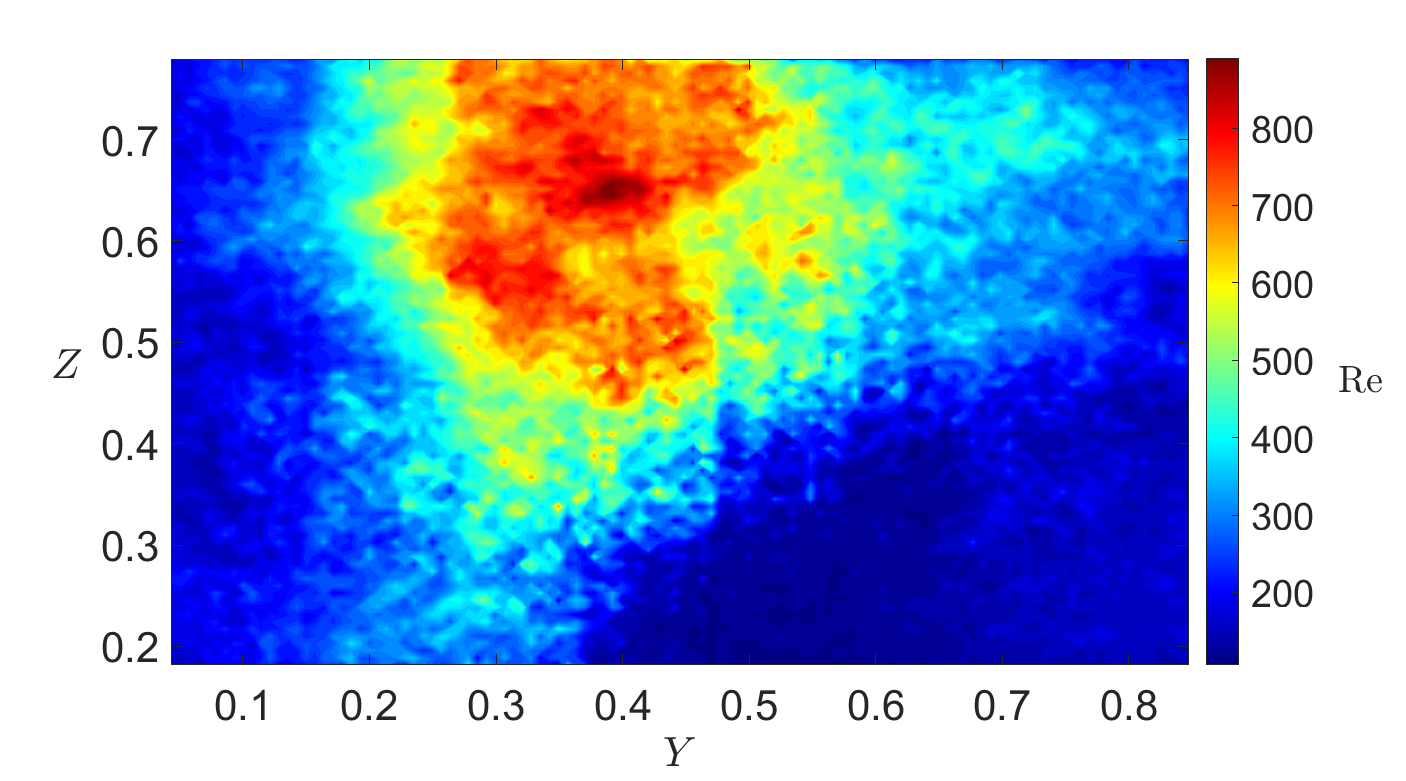}
\caption{\label{Fig11}
Distributions of the Reynolds number Re$=(u^{\rm (rms)}_y \, \ell_y + 2 u^{\rm (rms)}_z \, \ell_z)/\nu$
for convective turbulence
forced by one oscillating grid  (left panel) and by two oscillating grids (right panel).
The coordinates $Y$ and $Z$ are normalized by $L_z=26$ cm.
}
\end{figure*}

Distributions of various turbulent characteristics in the core flow of the chamber
have been obtained in the experiments with forced
convective turbulence (see Figs.~\ref{Fig4}--\ref{Fig11}).
In particular, in Fig.~\ref{Fig4} we show the distributions of the turbulent velocity
$|u^{\rm (rms)}| = \left[\langle u_y^2 \rangle + \langle u_z^2 \rangle\right]^{1/2}$.
In these experiments, the turbulent kinetic energy is produced by buoyancy in convection
and forcing produced by oscillating grids.
Large-scale velocity shear also contributes to the turbulence production rate.

\begin{figure*}[t!]
\centering
\includegraphics[width=8.5cm]{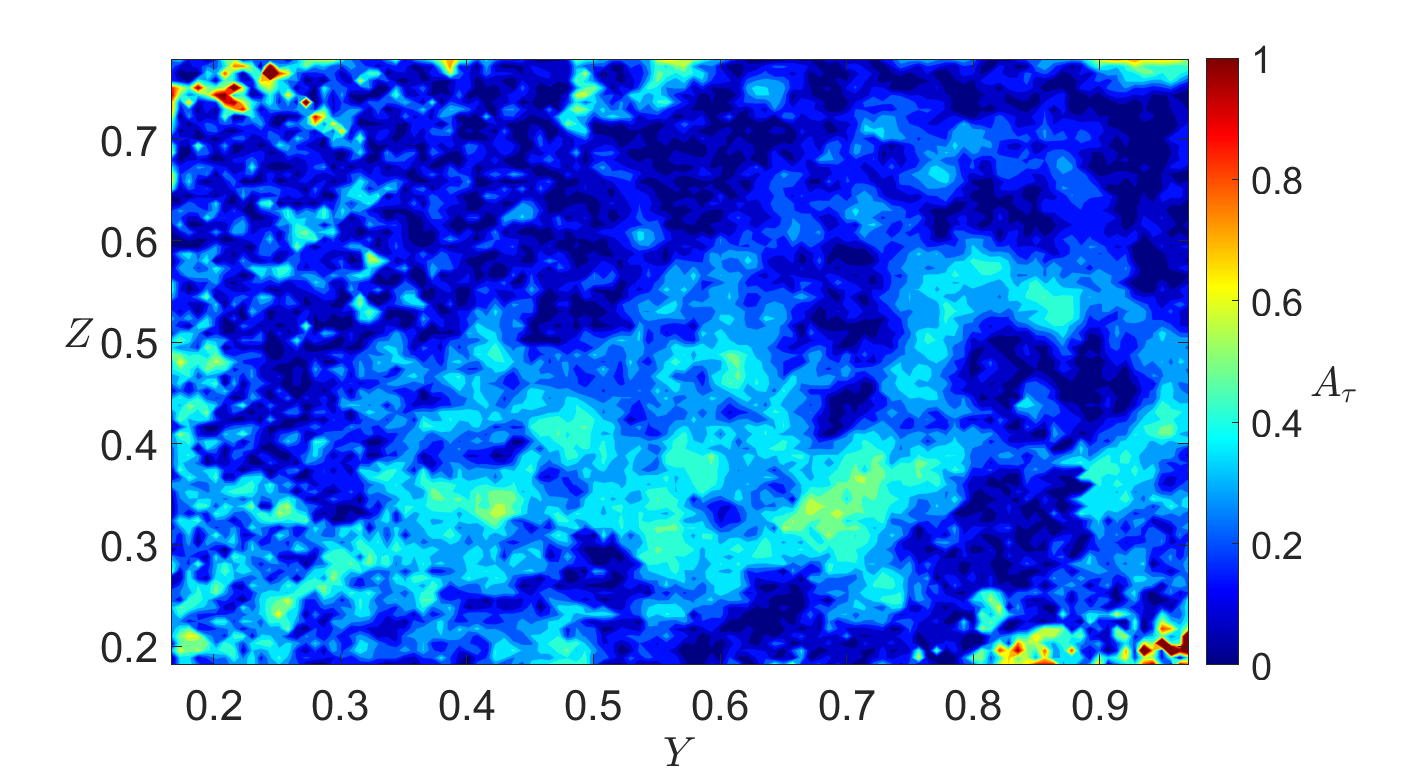}
\includegraphics[width=8.5cm]{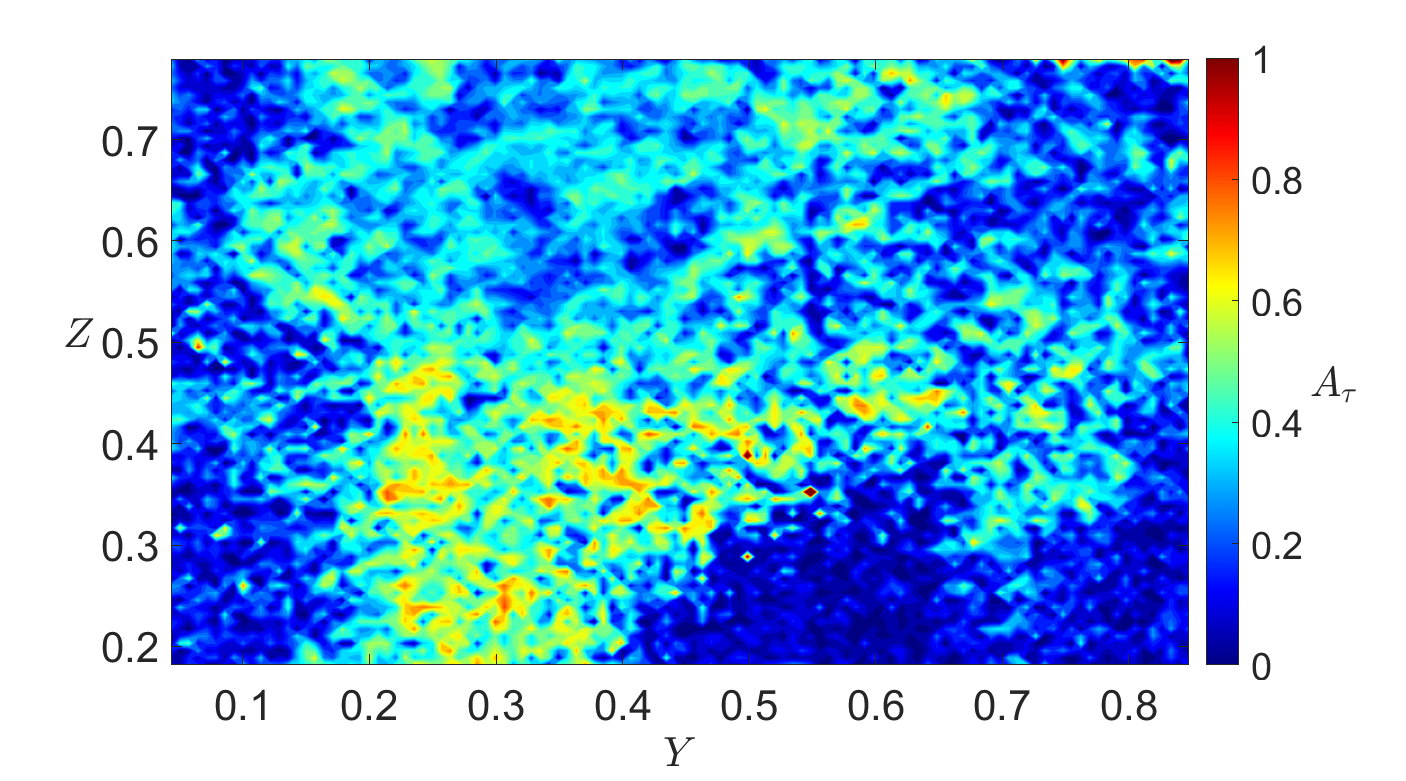}
\caption{\label{Fig12}
Distributions of the anisotropy of turbulent time $A_\tau =|\tau_z^{\rm (rms)} / \tau_y^{\rm (rms)} -1|$
for convective turbulence forced by one oscillating grid  (left panel) and by two oscillating grids (right panel).
The coordinates $Y$ and $Z$ are normalized by $L_z=26$ cm.
}
\end{figure*}

In Fig.~\ref{Fig5} we show the distributions of the ratio
of turbulent to mean velocities $u^{\rm (rms)} / |\meanUU|$ in convective turbulence.
For visualization of various turbulent regions,
the streamlines of the mean velocity field $\meanUU$ are also superimposed on this distribution.
In addition, the distributions of the turbulence anisotropy parameter $A_u=|u_z^{\rm (rms)} / u_y^{\rm (rms)} -1|$
are shown in Fig.~\ref{Fig6}.
It follows from Figs.~\ref{Fig4}--\ref{Fig6}, that velocity fluctuations are larger near oscillating grids
and turbulence intensity decreases with increase of the distance $Y$ from the grid.
For convective turbulence forced by two oscillating grids, turbulent velocity field is more uniform in the $Y$
direction and less anisotropic in comparison with that for one oscillating grid.

In Figs.~\ref{Fig7}--\ref{Fig10} we plot the
dependencies of  the horizontal and vertical turbulent velocities $u_y^{\rm (rms)}(Y)$ and $u_z^{\rm (rms)}(Y)$
and the horizontal and vertical integral turbulence scales $\ell_y(Y)$ and $\ell_z(Y)$
on the horizontal coordinate $Y$ in the core flow.
In various experiments with forced convection, we observe
the scalings $\ell_i(Y) \propto Y$ in the left and right ranges of the horizontal coordinate $Y$.
These scalings resemble qualitatively the results
of the early laboratory experiments conducted with one oscillating grid in isothermal turbulence  \cite{turn68,turn73,tho75,hop76,kit97,san98,med01}
and forced convective turbulence \citep{EKRL22,EKRL23,ZEKRL25},
where the integral turbulence length scale increases linearly with the distance $Y$ from a grid.

The measured components of the turbulent velocity and the integral turbulence scales
allow us to estimate the Reynolds number based on turbulent characteristics,
Re$=(u^{\rm rms}_y \, \ell_y + 2 u^{\rm rms}_z \, \ell_z)/\nu$
(see Fig.~\ref{Fig11}).
This expression for the Reynolds number estimate is obtained using the following arguments.
The Reynolds number is estimated as Re$=\tau_0 \langle {\bm u}^2  \rangle/\nu$,
where $\langle {\bm u}^2  \rangle= \langle u_x^2  \rangle + \langle u_y^2  \rangle+\langle u_z^2  \rangle$
and $\tau_0$ is the turbulent time in the integral scale.
We assume that the turbulent time is the same along $X$, $Y$ and $Z$ directions,
i.e., $\tau_0 = \ell_x/u^{\rm rms}_x = \ell_y/u^{\rm rms}_y = \ell_z/u^{\rm rms}_z$ and $u^{\rm rms}_x  \approx u^{\rm rms}_z$.
This yields the above estimate for the Reynolds number.

\begin{figure*}[t!]
\centering
\includegraphics[width=8.5cm]{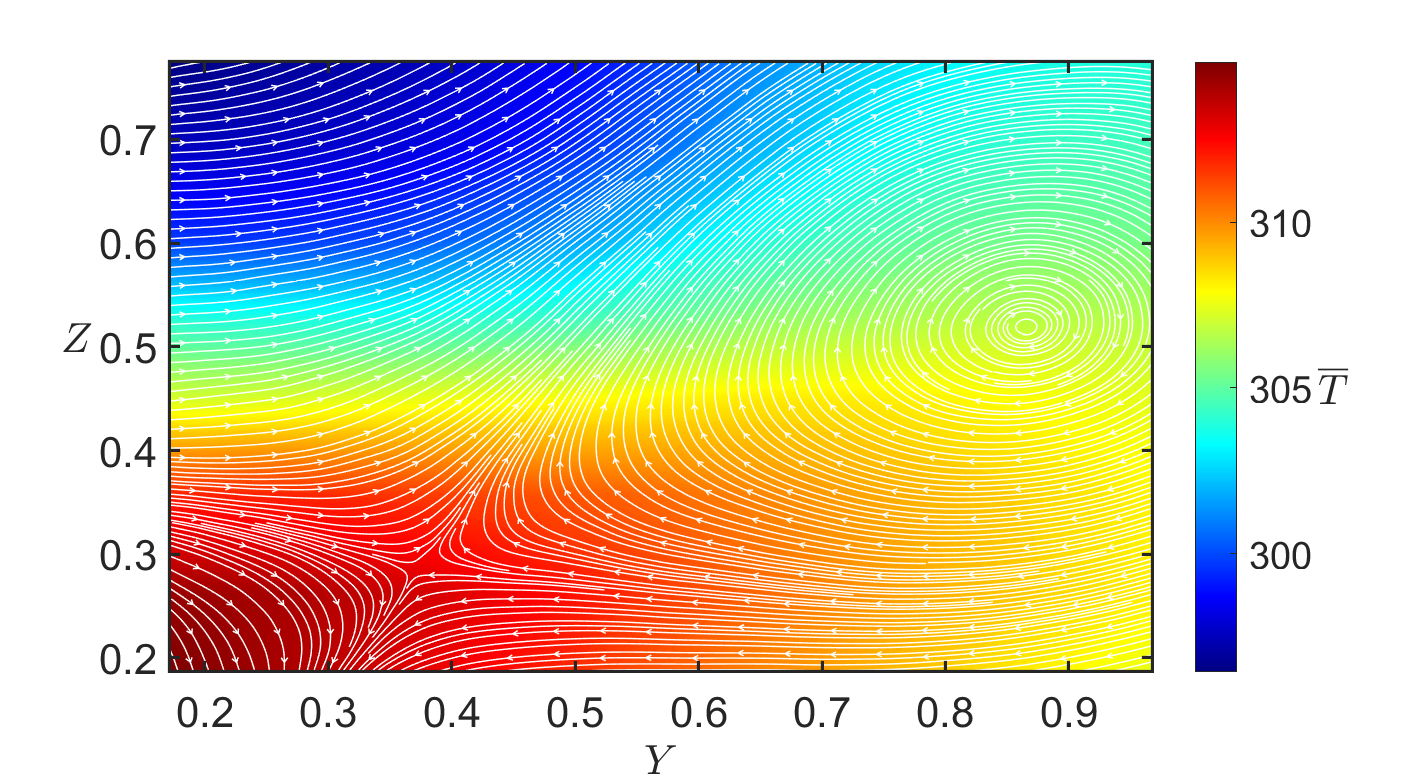}
\includegraphics[width=8.5cm]{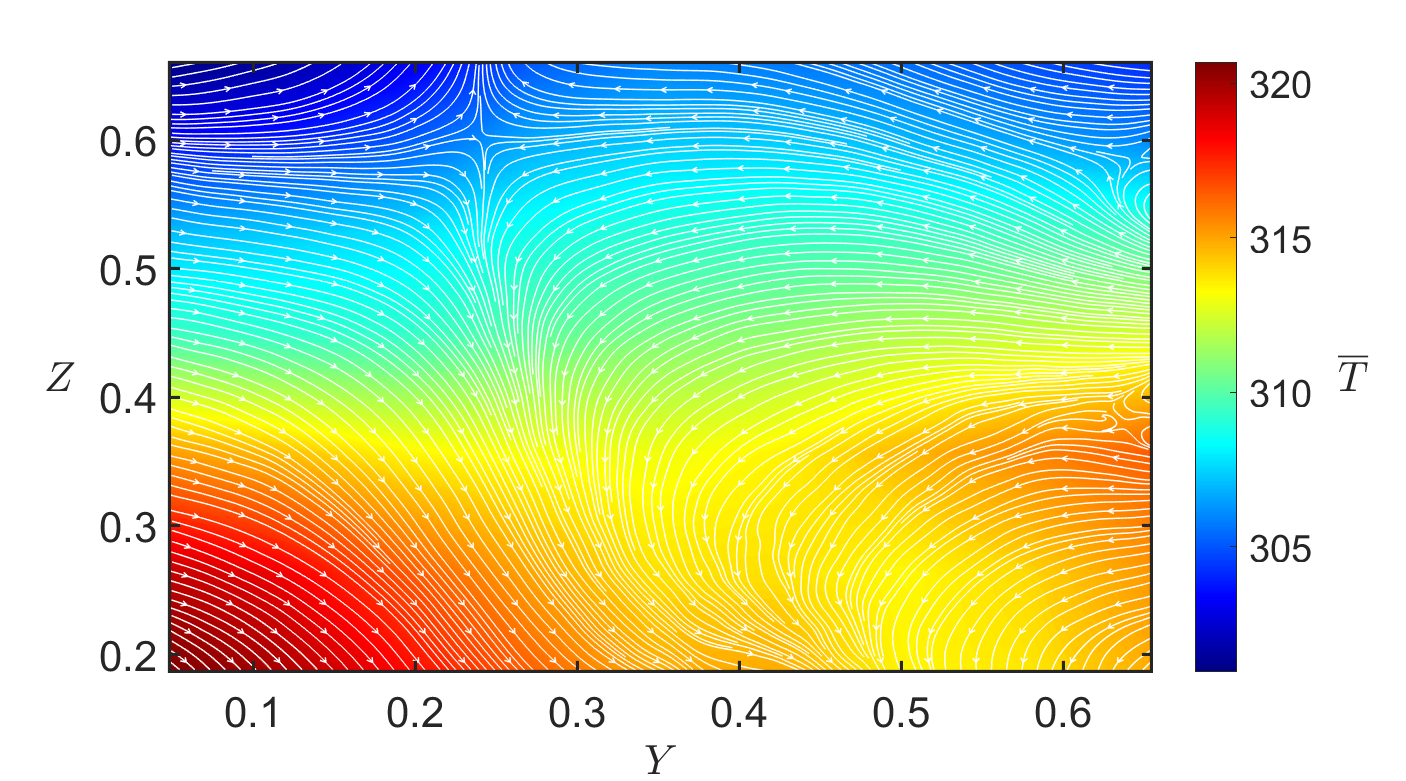}
\caption{\label{Fig13}
Distributions of the mean temperature $\meanT(Y,Z)$
for convective turbulence
forced by one oscillating grid  (left panel) and by two oscillating grids (right panel).
The streamlines (white)  of the mean velocity $\meanU$ are also superimposed on the temperature distribution.
The coordinates $Y$ and $Z$ are normalized by $L_z=26$ cm.
The mean temperature $\meanT(Y,Z)$ is measured in K.
}
\end{figure*}

\begin{figure*}[t!]
\centering
\includegraphics[width=8.5cm]{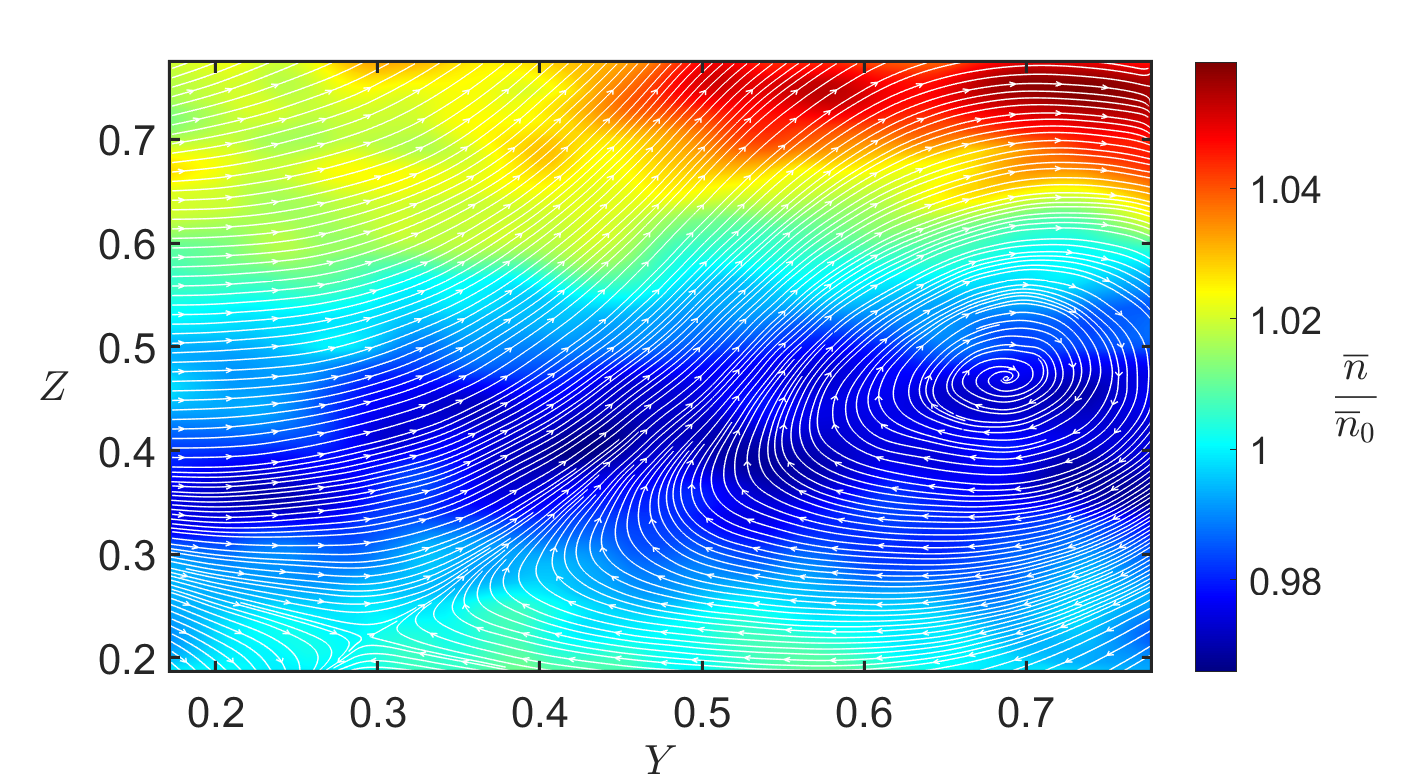}
\includegraphics[width=8.5cm]{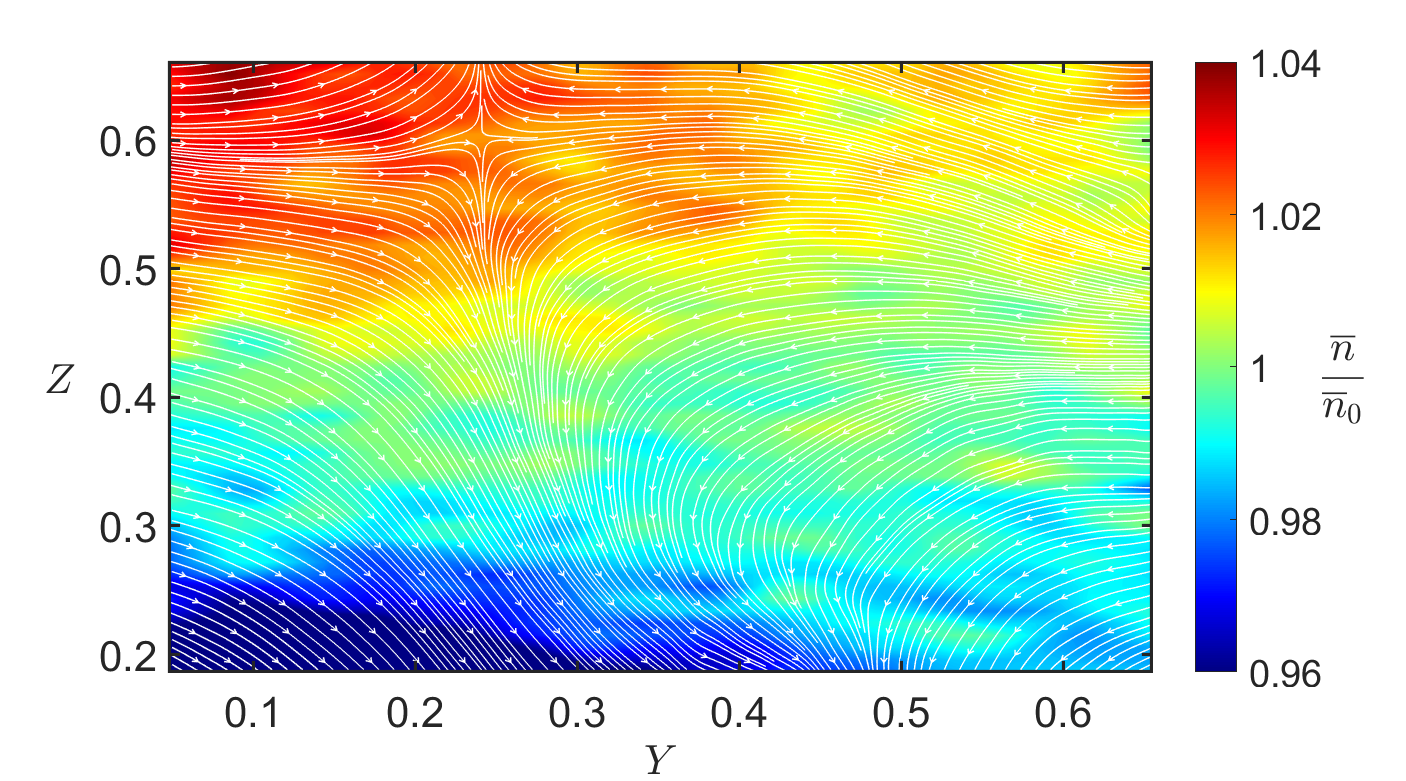}
\caption{\label{Fig14}
Distributions of the normalized mean particle number density $\meanN(Y,Z) / \meanN_0$
for convective turbulence forced by one oscillating grid  (left panel) and by two oscillating grids (right panel).
The streamlines (white)  of the mean velocity $\meanU$ are also superimposed on the particle distribution.
The coordinates $Y$ and $Z$ are normalized by $L_z=26$ cm.
}
\end{figure*}

The flow in the chamber is inhomogeneous and anisotropic, and
the assumption of identical turbulent time scales in all directions is an approximation.
In Fig.~\ref{Fig12} we show distributions of the anisotropy of turbulent time
$A_\tau =|\tau_z^{\rm (rms)} / \tau_y^{\rm (rms)} -1|$,
where $\tau_z^{\rm (rms)}=\ell_z/u_z^{\rm (rms)}$ and
$\tau_y^{\rm (rms)}=\ell_y/u_y^{\rm (rms)}$.
Since $A_\tau$ is small, we can use this assumption to estimate the Reynolds number.
On the other hand, the PIV system used in the present experiments provides two velocity components
($u^{\rm rms}_y$ and $u^{\rm rms}_z$) in the $YZ$ plane.
The third component along the $X$ axis is not directly measured.
Due to this reason, the turbulent quantities based on the measured velocity field
should be interpreted as estimates rather than exact values.
For the estimation of the Reynolds number based on experimentally accessible turbulent quantities,
we assume that $u^{\rm (rms)}_x \approx u^{\rm (rms)}_z$.
This assumption is supported by our previous experiments in a similar experimental set-up
with the oscillating grid turbulence
(see Figs.~4-5 in Ref.~\cite{BEE04}).
However, this assumption is not used for the evaluation of the accumulation of inertial particles.
In particular, the main analysis of turbulent transport of inertial particles is performed using
directly measured quantities in the vertical direction, including the vertical turbulent Reynolds
number and the vertical turbulent diffusion coefficient, which are
the physically relevant parameters governing the vertical turbulent transport of inertial particles and accumulation
process in our experiments (see below).

To investigate the phenomenon of turbulent thermal diffusion
in a forced convective turbulence, we measure the spatial distributions
of the mean temperature and the mean particle number density.
The initial spatial distributions of particles injected into the chamber,
are nearly homogeneous and isotropic.
In a convective turbulence, the spatial distributions of the mean particle number density
of inertial particles is expected to be inhomogeneous due to the effective drift velocity
${\bm V}^{\rm eff} = - \alpha D_T \, {\bm \nabla} \ln \overline{T}$, that
causes accumulation of inertial particles in the vicinity of the mean temperature minimum.

In Fig.~\ref{Fig13}, we show the spatial distributions of  the mean temperature $\meanT(Y,Z)$
in the core flow for turbulent convection forced by one oscillating grid  (left panel) and by two oscillating grids (right panel).
In Fig.~\ref{Fig14}, we also show the spatial distributions of the
normalized mean particle number density $\meanN(Y,Z)/\meanN_0$.
The mean particle number density $\meanN(Y,Z)$ in every point
in the experiments with forced convective turbulence
is normalized by
(i) the mean particle number density averaged over
the vertical column, where all area in the horizonal direction is divided into 65 vertical columns;
and (ii) the mean particle number density determined in the experiments with the isothermal
turbulence.

In spite of the presence of the  large-scale complicated fluid flows superimposed on the convective turbulence,
Figs.~\ref{Fig13}--\ref{Fig14} demonstrate the tendency of the localization of the maximum
of the mean particle number density in the vicinity of the minimum in the mean temperature, and vice versa.
In particular, particles are accumulated in the upper part of the chamber where
the mean temperature field $\meanT(Y,Z)$ achieves the minimum values.
As follows from Fig.~\ref{Fig14}, the maximum mean number density of inertial particles
is observed in the right upper part of the chamber for turbulent convection
forced by one oscillating grid, and in the left upper part of the chamber for turbulent convection
forced by two oscillating grids.

This occurs for the following reasons.
The Reynolds number reaches maximum values
in the right part of the chamber for turbulent convection
forced by one oscillating grid.
May be the latter is related to the effect of the mean flow
that is not completely destroyed by the forcing.
On the other hand, the Reynolds number is maximum
in the left part of the chamber for turbulent convection
forced by two oscillating grids (see Fig.~\ref{Fig11}).
In the regions with maximum Reynolds number,
the turbulent diffusion coefficient $D_T$ reaches the maximum values.
Since  the effective drift velocity ${\bm V}^{\rm eff}$ of inertial particles is proportional to $D_T$,
it also reaches the maximum values in these regions.
This explains why in these regions the mean number density of inertial particles reaches maximum values.
The above conclusions are valid in the turbulent regions with large mean temperature gradients.
For these regions, the dominant effect of large-scale particle clustering
is turbulent thermal diffusion.
Deviations from this trend occurs in the regions with strong mean fluid motions
where the mean temperature gradient is small.

\begin{figure*}[t!]
\centering
\includegraphics[width=8.5cm]{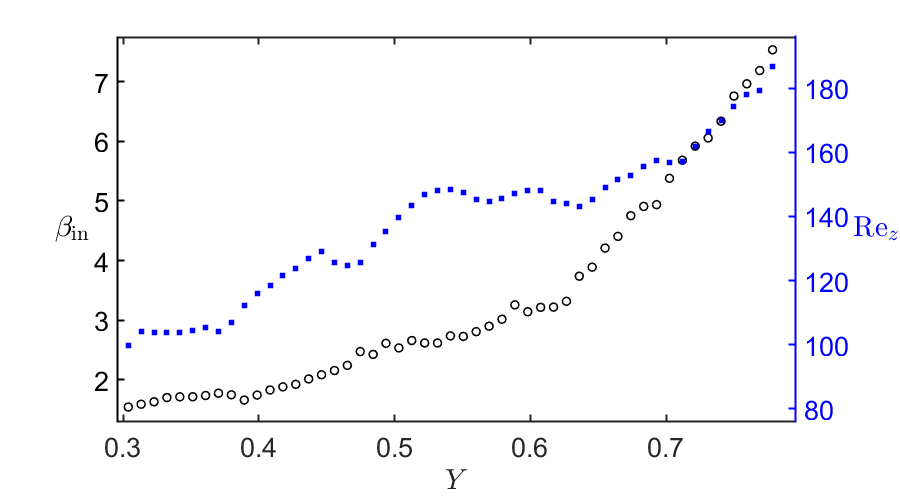}
\includegraphics[width=8.5cm]{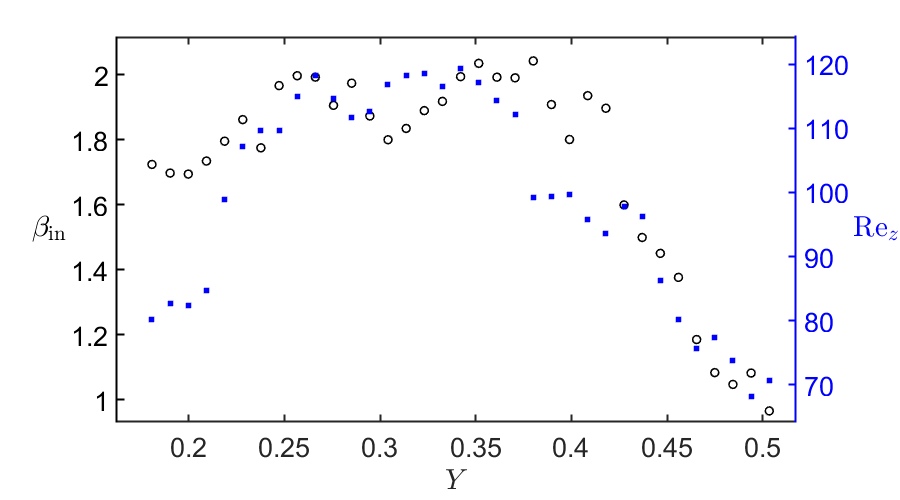}
\caption{\label{Fig15}
Dependencies of the parameter $\beta_{\rm in}$ (circles, grey) for inertial particles
and the vertical Reynolds number ${\rm Re}_z= u^{\rm rms}_z \, \ell_z/\nu$
(squares, blue) on the horizontal coordinate $Y$
for convective turbulence forced by one oscillating grid  (left panel) and by two oscillating grids (right panel).
The coordinate $Y$ is normalized by $L_z=26$ cm.
}
\end{figure*}

\begin{figure*}[t!]
\centering
\includegraphics[width=8.5cm]{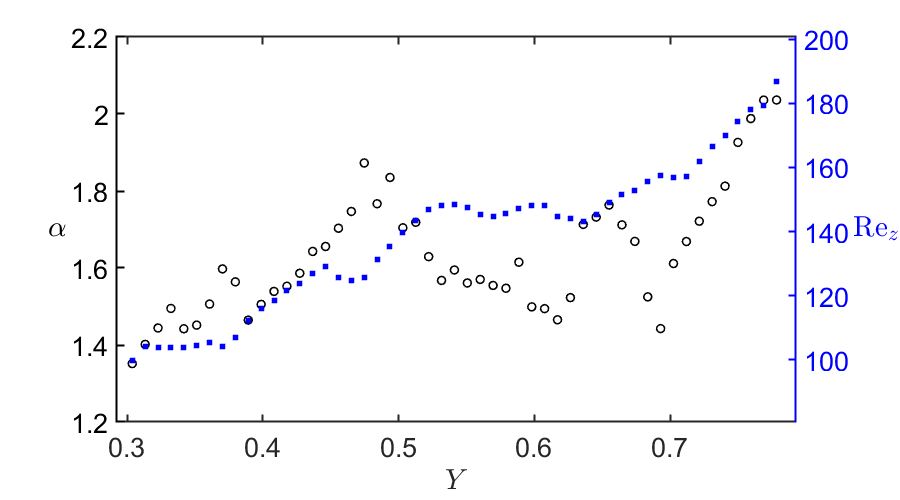}
\includegraphics[width=8.5cm]{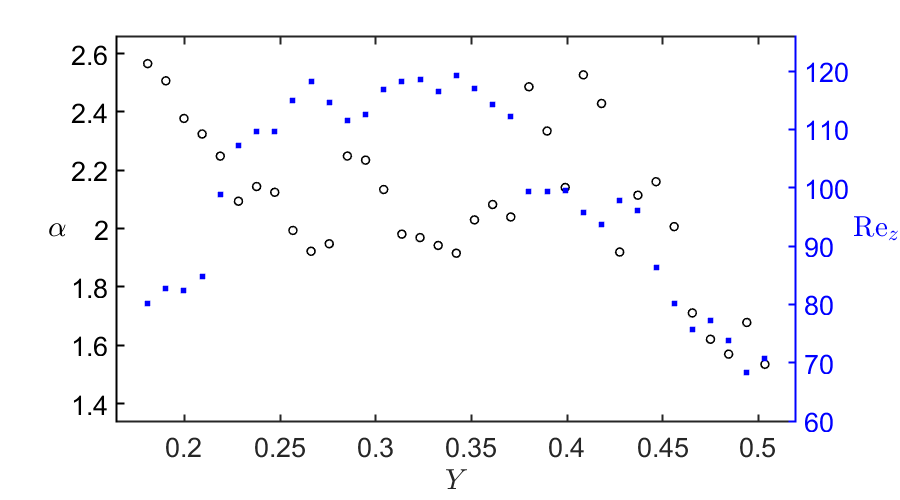}
\caption{\label{Fig16}
Dependencies of the parameter $\alpha$ (circles, grey) for inertial particles
and the vertical Reynolds number ${\rm Re}_z= u^{\rm rms}_z \, \ell_z/\nu$
(squares, blue) on the horizontal coordinate $Y$
for convective turbulence forced by one oscillating grid  (left panel) and by two oscillating grids (right panel).
The coordinate $Y$ is normalized by $L_z=26$ cm.
}
\end{figure*}

As can be seen in Fig.~\ref{Fig13}, the mean temperature gradients in the vertical direction
are significantly stronger than those in the horizontal directions.
In contrast, horizontal gradients of mean temperature are more localized and comparatively weak.
Therefore, the dominant contribution to turbulent thermal diffusion is associated
with the vertical temperature gradient, which justifies the use of a one-dimensional vertical formulation
for particle turbulent transport [see the steady-state solution~(\ref{A2}) for the equation for the mean
particle number density].

For quantitative analysis of the experimental results,
let us use Eq.~(\ref{A2}) for $\meanN(Z)$.
Taking into account that $\meanT(Z) = \meanT_{\rm b} + \delta \meanT$, and expanding the function
$(1 + \delta \meanT/\meanT_{\rm b})^{-\alpha} \approx 1 - \alpha (\delta \meanT/\meanT_{\rm b})$
in the Taylor series for $|\delta \meanT| \ll \meanT_{\rm b}$, we obtain that
\begin{eqnarray}
{\meanN(Z) \over \meanN_{\rm b}} \approx  - \beta {\biggl(\meanT(Z) - \meanT_{\rm b}\biggr) \over \meanT_{\rm b}}   ,
\label{A3}
\end{eqnarray}
where $\meanN_{\rm b} = \meanN(Z=0)$ and $\meanT_{\rm b}=\meanT(Z=0)$, and
the parameter $\beta$ is
\begin{eqnarray}
\beta = \alpha \exp \biggl(\int_0^Z {\meanU_z - V_{\rm g} \over D^{^{\rm T}}_z} \,{\rm d} Z' \biggr) .
\label{A4}
\end{eqnarray}
Here  $D^{^{\rm T}}_z$ is the vertical turbulent diffusion coefficient.

We stress that the spatial variations of the mean temperature
considered in our analysis in Eq.~(\ref{A3}) are local
and do not represent the total gradient of the mean temperature
based on the temperature difference $\Delta T = 50$~K
between the bottom and upper walls of the chamber in the experimental setup.
As a result, the temperature difference $\delta \meanT$ used in these calculations
is much smaller than the boundary temperature $\meanT_{\rm b}$.

The ratio of the parameters $\beta$ for inertial particles $\beta_{\rm in}$ and noninertial particles $\beta_{\rm nonin}$
is given by
\begin{eqnarray}
{\beta_{\rm in} \over \beta_{\rm nonin}} = \alpha \exp \biggl(- V_{\rm g} \, \int_0^Z {{\rm d} Z'\over D^{^{\rm T}}_z}  \biggr) ,
\label{A5}
\end{eqnarray}
where we take into account that for noninertial particles the parameter $\alpha=1$.
Equation~(\ref{A5}) allows us to determine the parameter $\alpha$ for inertial particles
in the experiments with forced convective turbulence.
In our previous experimental studies of turbulent thermal diffusion \cite{EKRL23},
the parameter $\beta$ has been introduced for non-inertial particles,
where it directly reflects the balance between turbulent transport and temperature gradients.
In the present study, we extend this definition to inertial particles in a consistent empirical manner,
by extracting the slope of the relationship between the normalized particle number density and the normalized temperature field.
While this parameter does not have the same direct theoretical interpretation as it does for the noninertial particles,
it provides a useful measure of particle response to temperature gradients.
This allows us to determine the parameter $\alpha$ as
\begin{eqnarray}
\alpha =\left({\beta_{\rm in} \over \beta_{\rm nonin}} \right) \,   \exp \biggl(V_{\rm g} \, \int_0^Z {{\rm d} Z'\over D^{^{\rm T}}_z}
 \biggr) ,
\label{A55}
\end{eqnarray}
which does not depend explicitly on the mean velocity field, and it
quantifies the effect of particle inertia on turbulent thermal diffusion.

Using the measured spatial distributions of the normalised mean particle number density $\meanN(Y,Z)/\meanN_0$
and the mean temperature $\meanT(Y,Z)$,
we determine the parameter $\beta$ in every vertical column
for inertial particles (having the diameter $10 \mu m$) and
in separate experiments with the same flow and temperature conditions for
noninertial particles (having the diameter $0.7 \mu m$).
The entire domain along the horizontal direction is divided
in 65 vertical columns.

To determine the distribution of the mean  temperature field,
the temperature measurements are performed with a spatial resolution of 1 cm in the horizontal direction
and 2.16 cm in the vertical direction (see Sec.~\ref{sect3}).
The measured data are imported directly into MATLAB and displayed using contour plots.
No additional smoothing, filtering, or artificial data manipulation are applied.
The contour representation only relies on the standard graphical interpolation used by MATLAB
for contour visualization and does not modify the underlying measured temperature field.
The time averaging of the temperature field in every point yields the mean temperature field.

The fitting procedure is applied locally, by identifying vertical segments within each column
where a clear linear relationship between the normalized mean number density and the normalized mean temperature is observed.
This ensures that Eq.~(\ref{A3}) is used only in regions where its underlying assumptions are satisfied.
The absence of a single global linear relation across the entire domain is expected, as both the mean flow field
and the temperature gradient vary spatially.
These variations lead to changes in the local transport balance and, consequently, to spatial variations in the effective slope.
To quantify the validity of the linear relation, we evaluated the fit quality across all selected segments.
The results show consistently high agreement (with the coefficient of determination $R^2>0.9$
that measures how well a model explains the variability of the data).
This demonstrates that Eq.~(\ref{A3}) provides a robust local description of the experimental data.
This procedure yields the dependence of the parameter $\beta$ on the horizontal coordinate $Y$.

For illustration, in Fig.~\ref{Fig15} we plot the functions $\beta_{\rm in}(Y)$ for inertial particles
and the vertical Reynolds number ${\rm Re}_z(Y) = u^{\rm rms}_z \, \ell_z/\nu$
for convective turbulence forced by one and two oscillating grids.
The parameter $\beta_{\rm in}(Y)$ provides useful information regarding the local relationship
between the normalized particle number density and the temperature field,
and it characterises the contributions associated with turbulent transport of inertial particles
and the mean flow [see Eqs.~(\ref{A3})--(\ref{A55})].

The ratio of the parameter $\beta_{\rm in}(Y)$ obtained in the experiments with
inertial particles to the parameter $\beta_{\rm nonin}(Y)$ for noninertial particles yields the dependence of
the parameter $\alpha(Y)$ for inertial particles on the horizontal coordinate $Y$ according to Eq.~(\ref{A55}).
In Fig.~\ref{Fig16} we show the dependencies of the parameter $\alpha(Y)$ for inertial particles
and the vertical Reynolds number ${\rm Re}_z(Y)$ on the horizontal coordinate $Y$
for convective turbulence forced by one and two oscillating grids.
The parameter $\alpha(Y)$ characterises
the contribution associated with turbulent transport of inertial particles.

We emphasize that our analysis of turbulent thermal diffusion and the estimation
of the parameter $\alpha$ primarily relies on the dominant vertical gradients
in the present flow configuration.
For this reason, we use the vertical Reynolds number Re$_z$
and the vertical turbulent diffusion coefficient $D^{^{\rm T}}_z$
as more physically relevant quantities in the analysis.

For convective turbulence forced by one oscillating grid, the parameter $\alpha$ for inertial particles
increases with the vertical Reynolds number ${\rm Re}_z$ varying from $\alpha=1.2$
to $\alpha=2.4$.
For convective turbulence forced by two oscillating grids, the behavior of the parameter $\alpha$ for inertial particles
is more complicated, but it varies from $\alpha=1.6$
to $\alpha=2.6$.
These results imply that the effect of accumulation of inertial particles is stronger
than that for noninertial particles, which
is in agreement with the theoretical predictions \cite{RI21,EKR98}.
Note also that the effective drift velocity caused by turbulent thermal diffusion
is in the vertical direction, while the effective drift velocity due to turbophoresis
is in horizontal direction.
In particular, the horizontal gradients of the turbulence intensity
are much larger than that in the vertical direction.
In addition, for the conditions pertinent to our experiments, turbulent thermal diffusion
is more effective than turbophoresis.

In our experiments the mean fluid velocity as well as the turbulent velocity
are much larger than the terminal fall velocity ($V_{\rm g}$ is about 0.33 cm/s)
for inertial particles having the diameter $10 \mu m$.
Accordingly, the contribution of the settling velocity to the the parameter $\alpha$
is small and can be neglected to leading order.
We note also that while particles possess a finite downward settling velocity, our measurements reveal
that the resulting particle number density field exhibits an effective upward transport.
Moreover, the sampling time in each experiment is relatively short and repeated over
many independent realizations.
This allows us to construct statistically meaningful averages over an ensemble of
experiments with the same flow field, effectively capturing the particle distribution
during the interval between their dispersion in the flow and their eventual settling.
These considerations indicate that, although gravitational settling is present,
it does not dominate the observed accumulation patterns,
which are primarily governed by turbulent transport mechanisms.

The strength of turbulent thermal diffusion is expected
to increase with the temperature gradient and the Rayleigh number
by increasing the temperature difference $\Delta T$
between the bottom and upper walls of the chamber.
Such a tendency has been found in our previous
experimental study of turbulent convection and turbulent thermal diffusion
for noninertial particles, see Refs.~\cite{AEKR17,BEKR09,BELR20}.

The increase of the Reynolds number enhances both
the effective drift velocity of inertial particles caused by turbulent thermal diffusion and
the turbulent diffusion coefficient, see Refs.~\cite{EEKR04,EEKR11,EEKR13}.
Competition between turbulent thermal diffusion and
turbulent diffusion affects the large-scale number density distribution for inertial particles.
However, a full parametric experimental study is beyond the scope of the present work.

\section{Discussion and conclusions}
\label{sect5}

Turbulent thermal diffusion of inertial particles
characterising a non-diffusive contribution to turbulent
flux of particles, has been studied in laboratory experiments with convective turbulence
forced by one or two oscillating grids in the airflow.
Fluid velocity field and spatial distribution of
inertial particles have been measured by Particle Image Velocimetry system,
and temperature field is measured using a temperature probe equipped with 12
E - thermocouples.
The experiments have shown the formation of large-scale clusters of inertial particles
in the regions of minimum of the mean temperature due to
effective drift velocity directed opposite to the gradient
of the mean fluid temperature.
Our experiments show that the effect of accumulation of inertial particles
(having the diameter $10 \mu m$) is stronger
than that for noninertial particles (having the diameter $0.7 \mu m$).
In particular, the measured maximum value of the parameter $\alpha$ for inertial particles
characterising the large-scale particle clustering reaches $\alpha_{\rm max} \approx 2.5$,
while for noninertial particles $\alpha=1$.
Therefore, as follows from our experiments,  the effective drift velocity due to turbulent thermal diffusion
is in 1.5 -- 2.5 times larger than that for noninertial particles depending on the level of turbulence.
For the conditions pertinent to our experiments, the effective drift velocity caused by turbulent thermal diffusion
is much stronger than that due to the turbophoresis.
This is in agreement with the theoretical predictions \cite{RI21,EKR98}.

The observed particle accumulation is primarily a bulk effect caused by turbulent thermal diffusion
and associated with the temperature gradient within the flow, rather than being driven by boundary effects.
This conclusion is supported by the spatial distribution of particle concentration (see Fig.~\ref{Fig14}),
which does not show localization near boundaries
but rather follows the large-scale temperature structure of the flow.

We also note that two heat exchangers with rectangular pins $3 \times 3 \times 15$
mm are attached to the bottom and top
walls of the chamber, which allow us to form a large vertical mean temperature gradient
in the core of the fluid flow.
This is the reason why near the top and bottom walls of the chamber,
the vertical mean temperature gradient is not large, in contrast to
experiments with smooth surface boundaries.
This implies that possible thermophoretic effects or near-wall boundary-layer effects in
the vicinity of the heated/cooled walls are very weak
in comparison with the dominant turbulent thermal diffusion effect in the core of the fluid flow.

Our experimental results have been compared with trends reported in previous studies on turbulent thermal diffusion, including our earlier laboratory experiments with noninerial particles in oscillating-grid turbulence \cite{BEE04,EEKR04,EEKR06a,AEKR17,EKRL22,EKRL23} and multi-fan generated turbulence \cite{EEKR06b}
as well as studies using turbulence generators with an oscillating membrane and a steady grid \cite{ZEKRL25}
under both, unstable and stable stratifications.
In addition, comparisons are made with DNS results for
both, inertial and noninertial particles in stably-stratified turbulence  \cite{HKRB12,RKB18}.
In particular, the observed accumulation of particles near the minimum of the mean temperature
is consistent with previously reported behavior in all experimental and numerical studies.

In this experimental study of turbulent thermal diffusion, we use forced convection, where
the convective turbulence is produced by buoyancy due to
the temperature difference $\Delta T$
between the bottom and upper walls of the chamber.
Forcing by the oscillating grids allows to destroy the large-scale circulation
and decrease mean velocity field in comparison with velocity fluctuations.
The turbulent kinetic energy is produced by
buoyancy and forcing caused by oscillating grids.
Increasing the oscillating frequency of the grids increases Reynolds number.
This enhances both, the turbulent diffusion coefficient and the effective drift velocity of inertial particles
caused by turbulent thermal diffusion.
In addition, the increase of the Reynolds number, enhances the parameter $\alpha$.
As a result it affects the spatial distribution of the mean particle number density,
i.e., it affects large-scale particle accumulation patterns.
\\

\bigskip
\noindent
{\bf ACKNOWLEDGEMENTS}
\medskip

We are thankful to the referees for providing constructive comments
which improved our paper.
\\

\bigskip
\noindent
{\bf AUTHOR DECLARATIONS}

\medskip
{\bf  Conflict of Interest}

The authors have no conflicts to disclose.
\\

\bigskip
\noindent
{\bf DATA AVAILABILITY}
\medskip

The data that support the findings of this study are available from the corresponding author
upon reasonable request.

\end{document}